\documentclass[prb,twocolumn,floatfix,amsmath,amssymb]{revtex4-1}
\pdfoutput=1
\usepackage{times}
\usepackage{bbold}
\usepackage{graphicx}
\usepackage{dcolumn}
\usepackage{bm}
\usepackage{color}
\usepackage{hyperref}
\renewcommand{\vec}{\mathbf}

\newcommand{\me}{\mathrm{e}}
\newcommand{\mi}{\mathrm{i}}


\begin{document}

\title{Magnetic breakdown  and quantum oscillations in  electron-doped high temperature superconductor $\mathrm{Nd_{2-x}Ce_{x}CuO_{4}}$}

\author{Jonghyoun Eun}
\author{Sudip Chakravarty}
\affiliation{Department of Physics and Astronomy, University of
California Los Angeles, Los Angeles, California 90095-1547, USA}

\date{\today}

\begin{abstract}
Recent more precise experiments have revealed  both a slow and a fast 
quantum oscillation in the $c$-axis resistivity  of nearly optimal to overdoped 
electron-doped high temperature superconductor $\mathrm{Nd_{2-x}Ce_{x}CuO_{4}}$. Here we study this problem from the perspective of Fermi
surface reconstruction using an exact transfer matrix method and the Pichard-Landauer formula. In this method, neither quasiclassical approximations for magnetic breakdown, nor {\em ad ho}c broadening of  Landau levels,  are necessary to study the high field quantum oscillations. The underlying Hamiltonian is a mean field Hamiltonian that incorporates a two-fold commensurate Fermi surface reconsruction. While the specific mean field considered is the $d$-density wave, similar results can also be obtained by a model of a  spin density wave, as was explicitly demonstrated earlier. The results are consistent with an interplay of magnetic breakdown across small gaps in the reconstructed Fermi surface and Shubnikov-de Haas oscillations.
\end{abstract}

\pacs{} 
\maketitle

\section{Introduction}
 Quantum oscillations were first discovered~\cite{Doiron-Leyraud:2007} in the Hall coefficient of   hole-doped high temperature superconductor  $\mathrm{YBa_{2}Cu_{3}O_{6+\delta}}$ (YBCO) at high magnetic fields between
$ 35 - 62 T$ in the underdoped regime close to 10\%. Since then a number of measurements, in even higher fields and with greater precision using a variety of measurement techniques  have confirmed the basic features of this experiment. However,
the precise mechanism  responsible for oscillations  has become controversial.~\cite{Riggs:2011}   Fermi surface reconstruction due to  a density wave order that could
arise if superconductivity is ``effectively destroyed'' by high magnetic fields have been focus of some attention.~\cite{Chakravarty:2008,*Chakravarty:2008b,*Dimov:2008,*Millis:2007,*Yao:2011}

In contrast, similar quantum oscillation measurements in the doping range $15-17\%$  in  $\mathrm{Nd_{2-x}Ce_{x}CuO_{4}}$ (NCCO)~\cite{Helm:2009} seem easier to interpret, as the magnetic field range $30-65 T$
is far above the upper critical field, which is less than $10 T$. This  clearly places the material in the ``normal'' state, a source of contention in measurements 
in YBCO; in NCCO the crystal  structure consists of a  single CuO plane per unit cell, and, in contrast to YBCO, there are no complicating
 chains, bilayers, ortho-II potential, stripes, etc.~\cite{Armitage:2009} Thus, it would appear to be ideal for gleaning the mechanism of quantum oscillations. On the other hand, disorder in NCCO  is significant. It is believed that well-ordered chain materials of  YBCO contain much less disorder by comparison.  
 
In a previous publication,~\cite{Eun:2010} we mentioned in passing that it is not possible   to understand the full picture in NCCO  without magnetic breakdown effects, since the gaps are expected to be very small in the relevant regime of the parameter space. However, in that preliminary work the breakdown phenomenon  was not addressed; instead we focused our attention to the effect of disorder. Since then recent measurements~\cite{Helm:2010,*Kartsovnik:2011} have indeed revealed magnetic breakdown in the range $16-17\%$ doping, almost  to the edge of the superconducting dome. Here we consider the same transfer matrix method used previously,~\cite{Eun:2010} but include third neighbor hopping of electrons on the square planar lattice, without  which many experimental aspects cannot be faithfully reproduced, including quantitative estimates of the oscillation frequencies and breakdown effects. The third neighbor hopping  makes the numerical transfer matrix calculation more intensive because of the enlarged size of the matrix, but we were able to overcome the technical challenge. In this paper we also  analyze  the $c$-axis resistivity and the absence of the electron pockets in the experimental regime.

\section{Hamiltonian}
The mean field Hamiltonian for $d$-density wave~\cite{Chakravarty:2001} (DDW) in real space, in terms of the
site-based fermion annihilation and creation operators $c_{\vec{i}}$
and  $c_{\vec{i}}^{\dagger}$, is
\begin{equation}\label{eq:hamiltonian}
    H_{DDW}=\sum_{\vec{i}}\epsilon_\vec{i}c_\vec{i}^\dag c_\vec{i}+\sum_{
    \vec{i},\vec{j}}t_{\vec{i},\vec{j}}~\mathrm{e}^{ia_{\vec{i},\vec{j}}}c_\vec{i}^\dag
    c_\vec{j}+h. c\\.,
\end{equation}
where the nearest neighbor hopping matrix elements include DDW gap $W_{0}$ and are
\begin{equation}\label{Eq:DDWhopping}
\begin{split}
    t_{\vec{i},\vec{i}+\hat{\vec{x}}}&=t+\frac{iW_0}{4}(-1)^{(n+m)},\\
    t_{\vec{i},\vec{i}+\hat{\vec{y}}}&=t-\frac{iW_0}{4}(-1)^{(n+m)},\\
\end{split}
\end{equation}
where $(n,m)$ are a pair of integers labeling a site:
$\vec{i}=n\hat{\vec{x}}+m\hat{\vec{y}}$; the lattice constant $a$ will be set to unity unless otherwise specified .
In this paper we also include
both next nearest hopping matrix element, $t'$, and third nearest neighbor hopping matrix element $t''$.  A constant perpendicular magnetic field $B$ is included
via the Peierls phase factor
$a_{\vec{i},\vec{j}}=\frac{
e}{\hbar c}\int_\vec{j}^\vec{i}\vec{A}\cdot\mathrm{d}\vec{l}$, where
$\vec{A}=(0,-Bx,0)$ is the vector potential in the Landau gauge. The band parameters are chosen to be
$t~=~0.38eV$, $t'~=~0.32t$, and $t''~=~0.5t'$.~\cite{Pavarini:2001} The chemical
potential $\mu$ is adjusted to achieve the required doping level and is given in
Table~\ref{table1}, so is the DDW gap $W_{0}$.
We assume that the 
on-site energy is $\delta$-correlated white noise defined by the
disorder average $\overline{\epsilon_\vec{i}}=0$ and
$\overline{\epsilon_\vec{i}\epsilon_\vec{j}}=V_{0}^{2}\delta_{\vec{i},\vec{j}}$.
 Disorder levels for each of the cases studied
are also given there in Table~\ref{table1}. We have seen previously that longer ranged correlated
disorder lead to very similar results.~\cite{Jia:2009}

\begin{figure}
\begin{center}
\includegraphics[width=\linewidth]{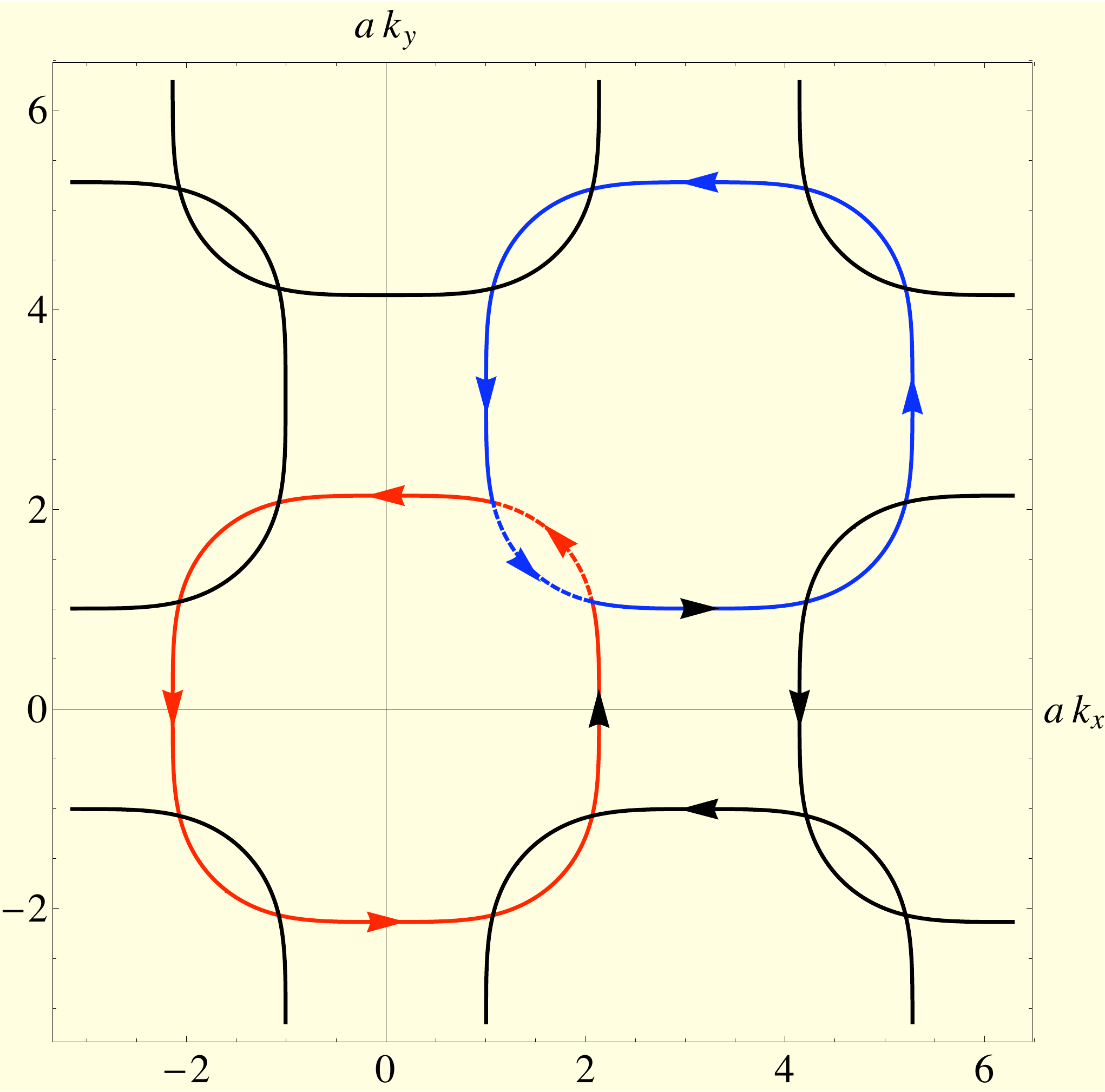}
\caption{(Color online) A  plot showing the breakdown junctions and electron trajectories across them in the extended Brillouin zone. The figure corresponds  NCCO with 17\% doping and  small DDW gap. Note that the reflection at the junctions involve a large change in momentum. The electron trajectories that lead to magnetic breakdown of small hole pockets are shown.} 
\label{fig:Three-Pocket}
\end{center}
\end{figure}

The Fermi surface areas (See Fig.~\ref{fig:Three-Pocket}) of the small hole pocket in the absence of disorder correspond to oscillation frequencies $330 T$ at 15\% doping,  $317 T$ at 16\% doping and $291 T$ at 17\% doping. These frequencies seem to be insensitive to $W_{0}$ within the range given in Table~\ref{table1}.

\section{The transfer matrix method}
Transfer matrix to compute the oscillations of the conductance is a powerful method. It requires 
neither quasiclassical approximation to investigate magnetic breakdown nor does it require {\em ad hoc} broadening of the
Landau level to incorporate the effect of disorder. Various models of disorder, both long and short-ranged, can be studied {\em ab initio}. The mean field Hamiltonian, being a quadratic non-interacting Hamiltonian, leads to a Schr\"odinger equation for the site amplitudes, which is then recast in the form of a transfer matrix; the full derivation is given in the Appendix. The conductance is then calculated by a formula that is well known in the area of mesoscopic physics, the Pichard-Landauer formula.~\cite{Pichard:1986,*Fisher:1981} This yields Shubnikov-de Haas oscillations of the $ab$-plane resistivity, $\rho_{ab}$. We show later how this can be related to the $c$-axis resistivity $\rho_{c}$ measured in experiments.

Consider a quasi-1D system, $N\gg M$, with a periodic boundary
condition along y-direction. Here $Na$ is the length in the $x$-direction and $Ma$ is the length in the $y$-direction, $a$ being the lattice spacing.  Let $\Psi_n = (\psi_{n,1},\psi_{n,2}, \ldots, \psi_{n,M})^T$, $n=1, \dots N$,
be the amplitudes on the slice $n$ for an
eigenstate with a given energy. Then the amplitudes on four
successive slices must satisfy the relation
\begin{widetext}
\begin{equation}
    \label{Eq:transfermatrix}
    \left[
    \begin{array}{c}
        \Psi_{n+2} \\
        \Psi_{n+1} \\
        \Psi_{n~~} \\
        \Psi_{n-1} \\
    \end{array}
    \right]
    =
    \left[
    \begin{array}{cccc}
        U_n^{-1} A_{n} & U_n^{-1} B_{n} & U_n^{-1} C_{n} & U_n^{-1} D_{n}\\
        1 & 0 & 0& 0\\
        0 & 1 & 0& 0\\
        0 & 0 & 1& 0\\
    \end{array}
    \right]
    \left[
    \begin{array}{c}
        \Psi_{n+1} \\
        \Psi_{n~~} \\
        \Psi_{n-1} \\
        \Psi_{n-2} \\
    \end{array}
    \right]
    =
    {\mathbf T}_{n}
    \left[
    \begin{array}{c}
        \Psi_{n+1} \\
        \Psi_{n~~} \\
        \Psi_{n-1} \\
        \Psi_{n-2} \\
    \end{array}
    \right]
\end{equation}
\end{widetext}
where $U_n$, $A_n$, $B_n$, $C_n$, $D_n$ are  $M\times M$ matrices.
The non-zero matrix elements of matrix $A_n$ are
\begin{eqnarray}
   (A_n)_{m,m}&=&-\left[-1-\frac{\mi W_0}{4}(-1)^{m+n}\right],\\
   (A_n)_{m,m+1}&=&-t'\me^{\mi(-n-\frac{1}{2})\phi},\\
   (A_n)_{m,m-1}&=&-t'\me^{\mi(n+\frac{1}{2})\phi},
\end{eqnarray}
where $\phi = Ba^2e/\hbar c$ is a constant. The elements of the matrix $B_n$ are 
\begin{eqnarray}
   (B_n)_{m,m}&=&\epsilon_{n,m}-\mu,\\
   (B_n)_{m,m+1}&=&\left[-1+\frac{\mi W_0}{4}(-1)^{m+n}\right]\me^{-\mi n\phi},\\
   (B_n)_{m,m-1}&=&\left[-1+\frac{\mi W_0}{4}(-1)^{m+n}\right]\me^{\mi n\phi},\\
   (B_n)_{m,m+2}&=&t^{\prime\prime}\me^{-\mi 2 n\phi},\\
   (B_n)_{m,m-2}&=&t^{\prime\prime}\me^{\mi 2 n\phi},
\end{eqnarray}
Here $C_n = A^{\dag}_n$ and
$D_n = - U_n = t^{\prime\prime} \mathbb{1}$, where $\mathbb{1}$ is the $M\times M$ identity matrix.

\begin{table}[htdp]
\caption{Parameters $W_{0}$ (DDW gap), $V_{0}$ (on-site disorder potential), and $\mu$ (chemical potential).}
\begin{center}\begin{tabular}{|ccccc|}
\hline
Figure & Gap $W_{0}$ (meV) & $V_{0}$ (disorder) & $\mu$ & doping (\%)\\
\hline\hline
Fig.~\ref{fig:gap5-1}& 5 & $0.2t$ & $0.057t$ & 17\\
Fig.~\ref{fig:gap5-2}& 5 & $0.4t$ & $0.057t$ & 17\\
Fig.~\ref{fig:gap5-3}& 5 & $0.6t$ & $0.057t$ & 17\\
Fig.~\ref{fig:gap10-1}& 10 & $0.2t$ & $0.057t$ & 17\\
Fig.~\ref{fig:gap10-2}& 10 & $0.4t$ & $0.057t$ & 17\\
Fig.~\ref{fig:gap10-3}& 10 & $0.6t$ & $0.057t$ & 17\\
Fig.~\ref{fig:gap15-1}& 15 & $0.2t$ & $0.0176t$ & 16 \\
Fig.~\ref{fig:gap15-2}& 15 & $0.4t$ & $0.0176t$ & 16 \\
Fig.~\ref{fig:gap15-3}& 15 & $0.6t$ & $0.0176t$ & 16 \\
Fig.~\ref{fig:gap30-1}& 30 & $0.2t$ & $0.0176t$ & 16\\
Fig.~\ref{fig:gap30-2}& 30 & $0.4t$ & $0.0176t$ & 16\\
Fig.~\ref{fig:gap30-3}& 30 & $0.6t$ & $0.0176t$ & 16\\
\hline
\end{tabular}
\end{center}
\label{table1}
\end{table}

The $4M$ Lyapunov exponents, $\gamma_{i}$, of
$\lim_{N\to\infty}({\cal T}_{N}{\cal T}_{N}^{\dagger})$, where
${\cal T}_{N}=\prod_{j=1}^{j=N}{\mathbf T}_{j}$, are defined by the
corresponding eigenvalues $\lambda_{i}=e^{\gamma_{i}}$.
 All the Lyapunov exponents
$\gamma_1>\gamma_2>\ldots>\gamma_{4M}$, are computed by a method
described in Ref.~\onlinecite{Kramer:1996}. However, the
matrix is not symplectic. Therefore all $4M$ eigenvalues are
computed. Remarkably,  except for a small
set, consisting of large eigenvalues, the rest of the eigenvalues do come in pairs
$(\lambda, 1/\lambda)$, as for the symplectic case, within our  numerical
accuracy. We have no analytical proof of this curious fact. Clearly,
large eigenvalues contribute insignificantly to the Pichard-Landauer~\cite{Pichard:1986}  formula
for the conductance, $\sigma_{ab}(B)$:
\begin{equation}
\sigma_{ab}(B) = \frac{e^{2}}{h} \text{Tr}\sum_{j=1}^{2M}\frac{2}{({\cal
T}_{N}{\cal T}_{N}^{\dagger})+({\cal T}_{N}{\cal
T}_{N}^{\dagger})^{-1}+2}.
\end{equation}
We have
chosen $M$ to be 32, smaller than our previous
work.~\cite{Eun:2010} The reason for this is that the matrix size including the third neighbor hopping  is larger $4M\times 4M$ instead of $2M\times 2M$. We
chose $N$  to be of the order of $10^{6}$, as before. This easily led to an accuracy better
than 5\% for the smallest Lyapunov exponent, $\gamma_{i}$, in all
cases.

\section{Magnetic breakdown and quantum oscillations}
We compute the conductance as a function of the magnetic field and then Fourier
transform the numerical data.   This procedure of course depends on the number of data points
sampled within a fixed range of the magnetic field, typically between $45 - 60 T$. As  the number
of sampling points increases,  the peaks become  narrower but greater in intensity, conserving the area under the peak. But the location of each peak and the
relative ratio of the intensities remain the same. In order to compare the Fourier transformed results, we keep the sampling points fixed in all cases to be 1200.

\begin{figure}
\begin{center}
\includegraphics[width=\linewidth]{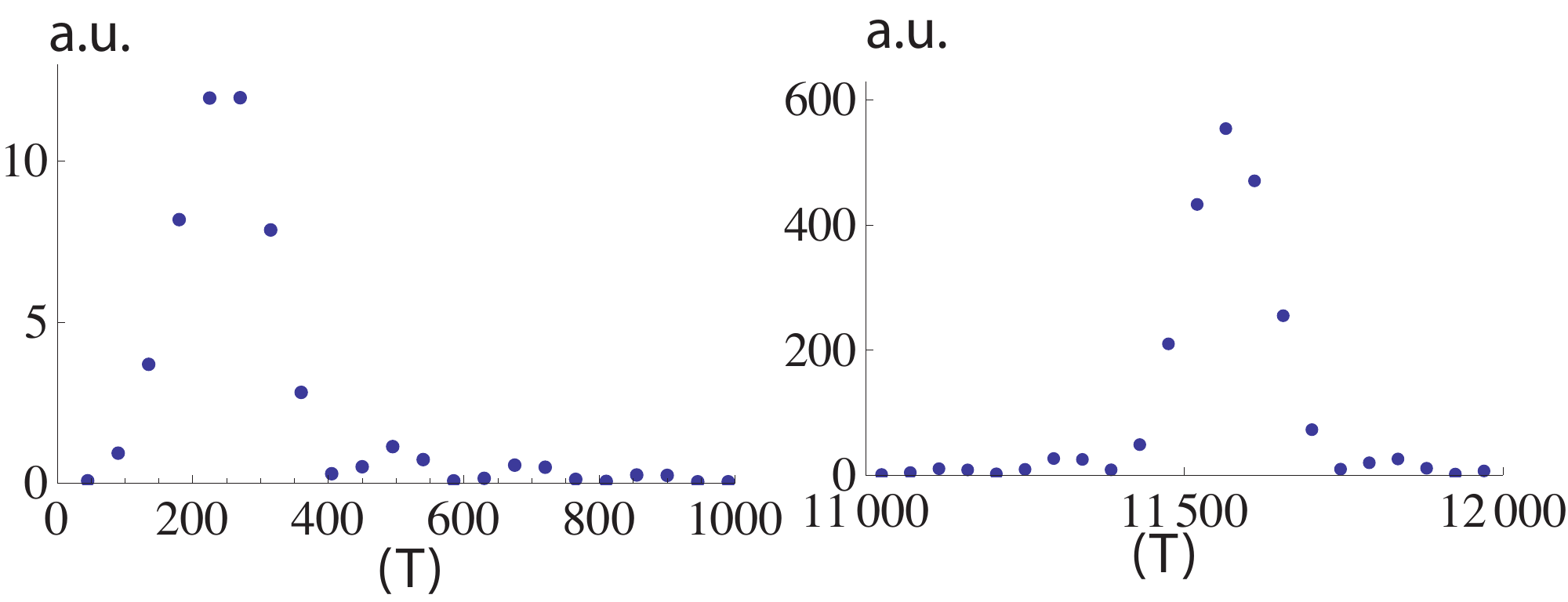}
\caption{(Color online) Fourier transform of the conductance oscillations with a smooth background term subtracted. The parameters correspond  to 17\%  doping  with a DDW gap of $ 5 meV$ and disorder $V_{0}=0.2 t$. The horizontal axis is in units of Tesla and the vertical axis is in arbitrary units.}
\label{fig:gap5-1}
\end{center}
\end{figure}

\begin{figure}
\begin{center}
\includegraphics[width=\linewidth]{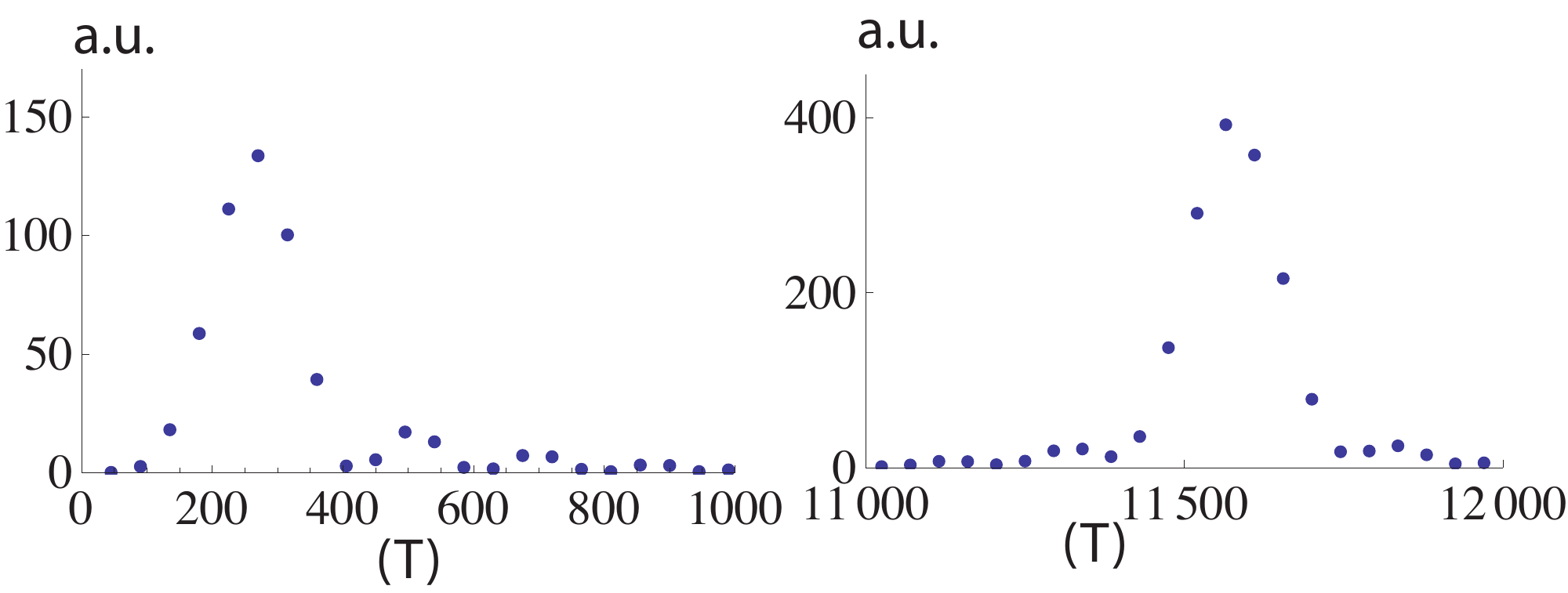}
\caption{(Color online) The same as in Fig.~\ref{fig:gap5-1} corresponding to 17\%  doping  but with a DDW gap of $ 5 meV$ and disorder $V_{0}=0.4 t$.}
\label{fig:gap5-2}
\end{center}
\end{figure}

\begin{figure}
\begin{center}
\includegraphics[width=\linewidth]{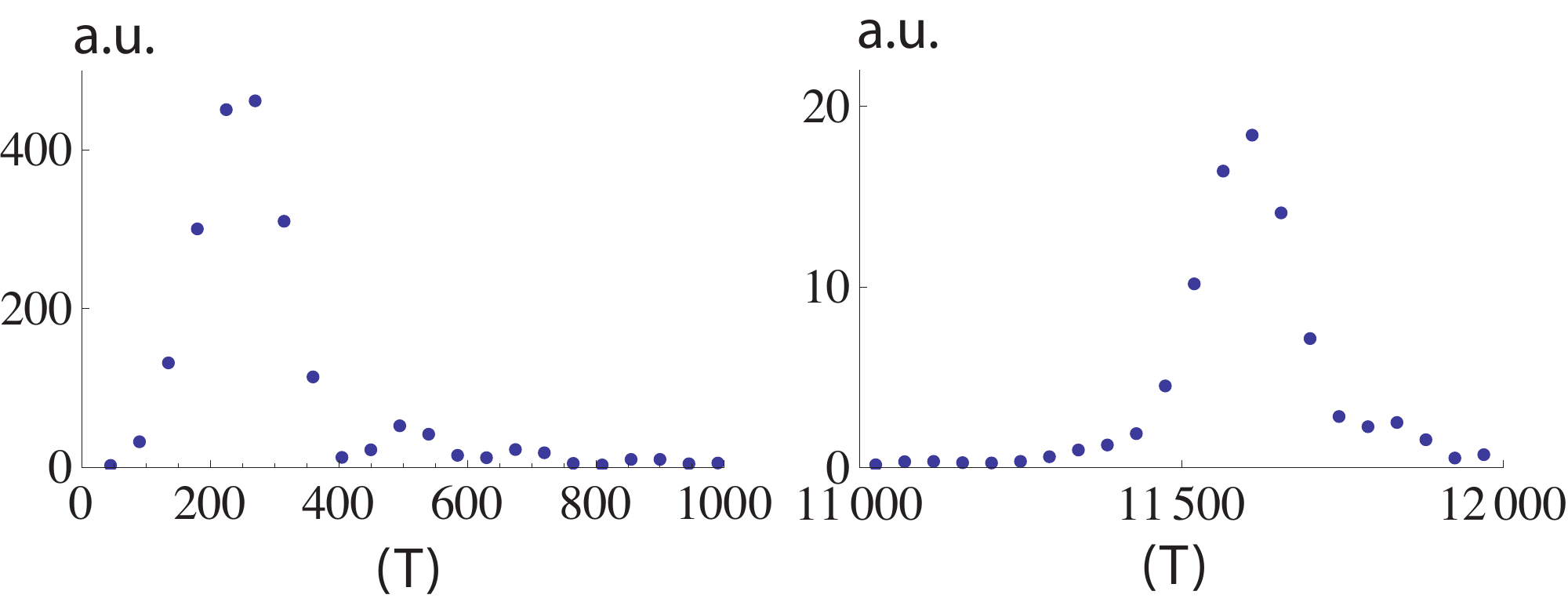}
\caption{(Color online) The same as in Fig.~\ref{fig:gap5-1} corresponding to 17\% doping but  with a DDW gap of $ 5 meV$ and disorder $V_{0}=0.6 t$.}
\label{fig:gap5-3}
\end{center}
\end{figure}

\begin{figure}
\begin{center}
\includegraphics[width=\linewidth]{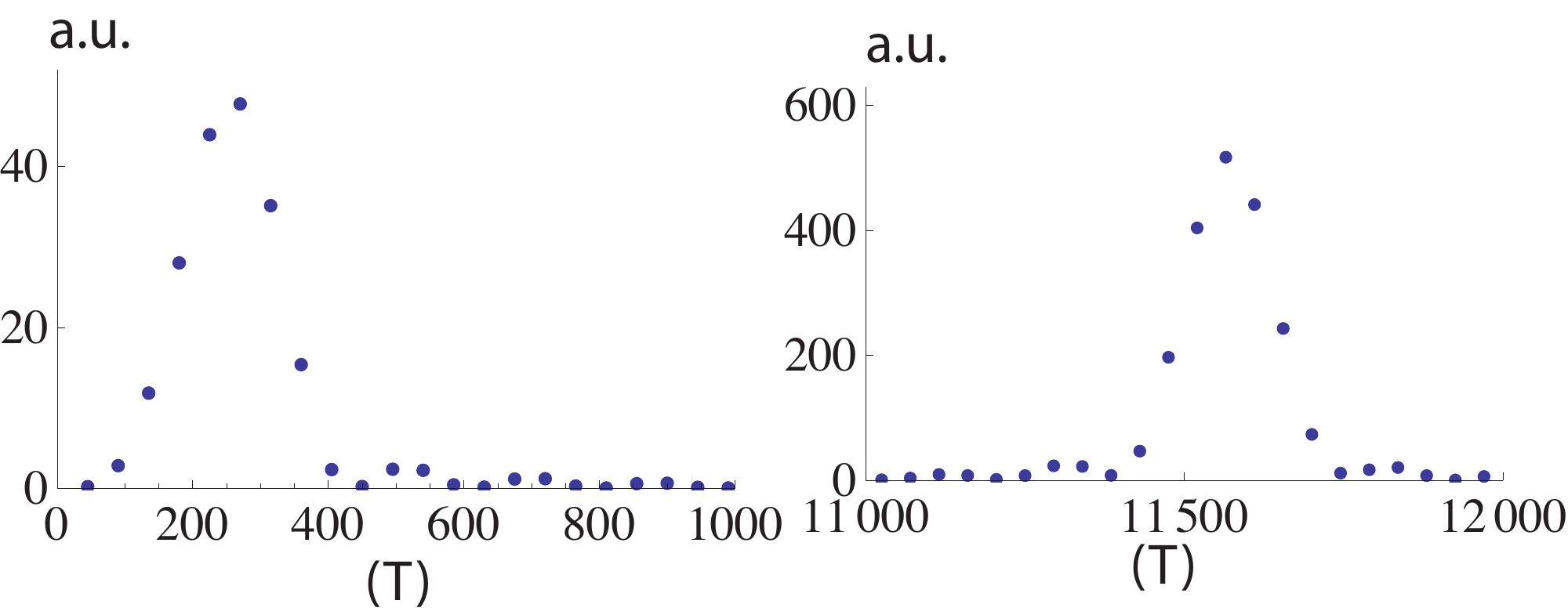}
\caption{(Color online) The same as in Fig.~\ref{fig:gap5-1} corresponding to 17\% doping but with a DDW gap of $ 10 meV$ and disorder $V_{0}=0.2 t$.}
\label{fig:gap10-1}
\end{center}
\end{figure}

\begin{figure}
\begin{center}
\includegraphics[width=\linewidth]{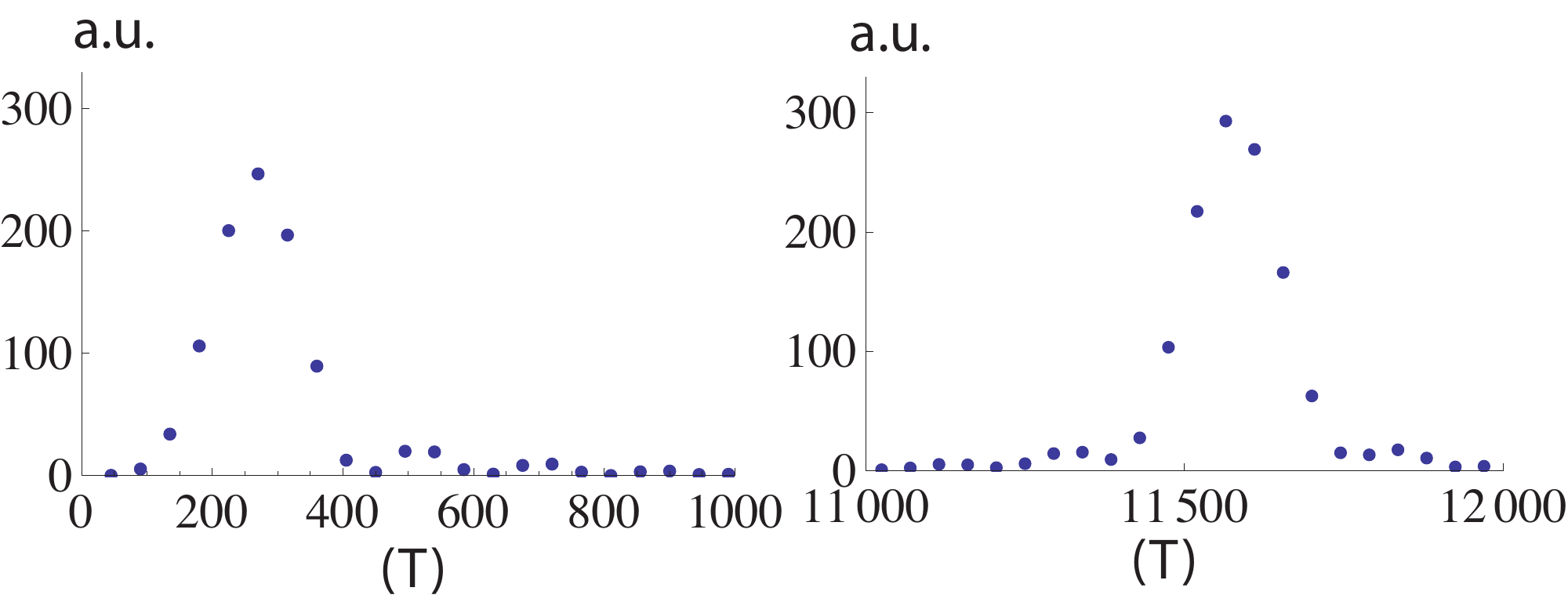}
\caption{(Color online) The same as in Fig.~\ref{fig:gap5-1} corresponding to 17\% doping but with a DDW gap of $ 10 meV$ and disorder $V_{0}=0.4 t$.}
\label{fig:gap10-2}
\end{center}
\end{figure}

\begin{figure}
\begin{center}
\includegraphics[width=\linewidth]{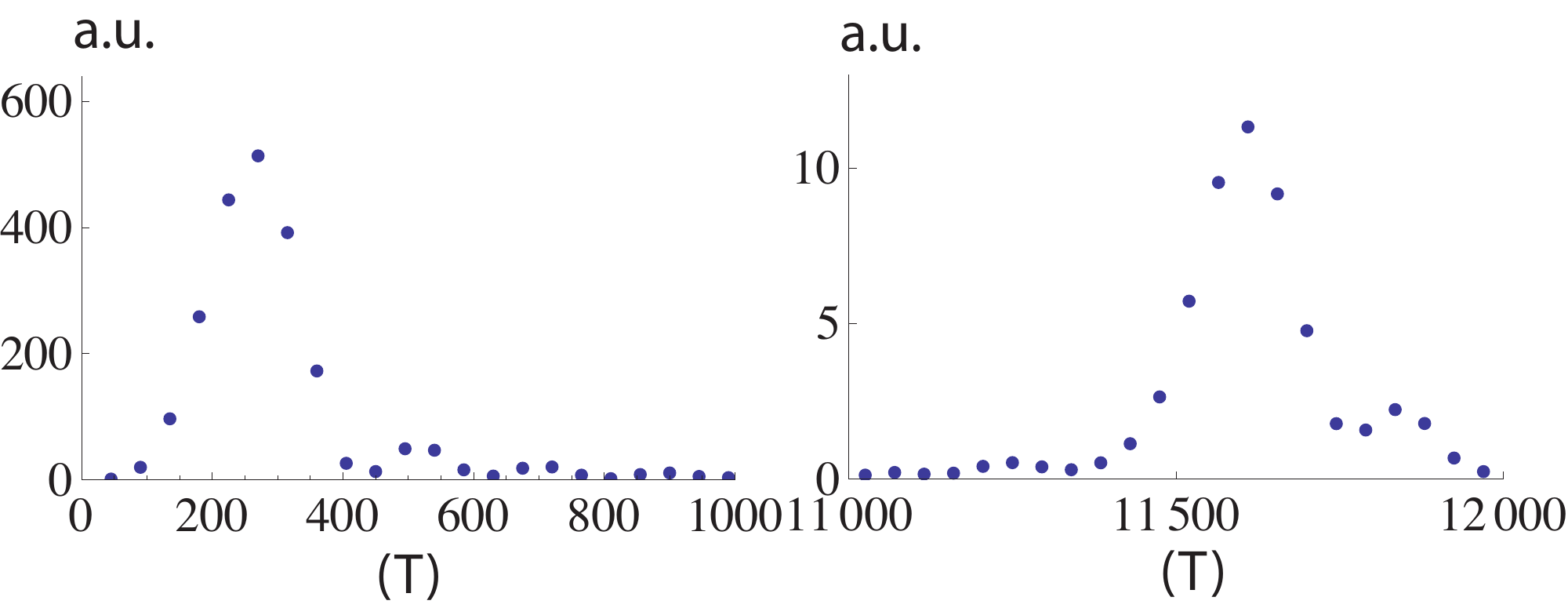}
\caption{(Color online) The same as in Fig.~\ref{fig:gap5-1} corresponding to 17\% doping but with a DDW gap of $ 10 meV$ and disorder $V_{0}=0.6 t$.}
\label{fig:gap10-3}
\end{center}
\end{figure}

\begin{figure}
\begin{center}
\includegraphics[width=\linewidth]{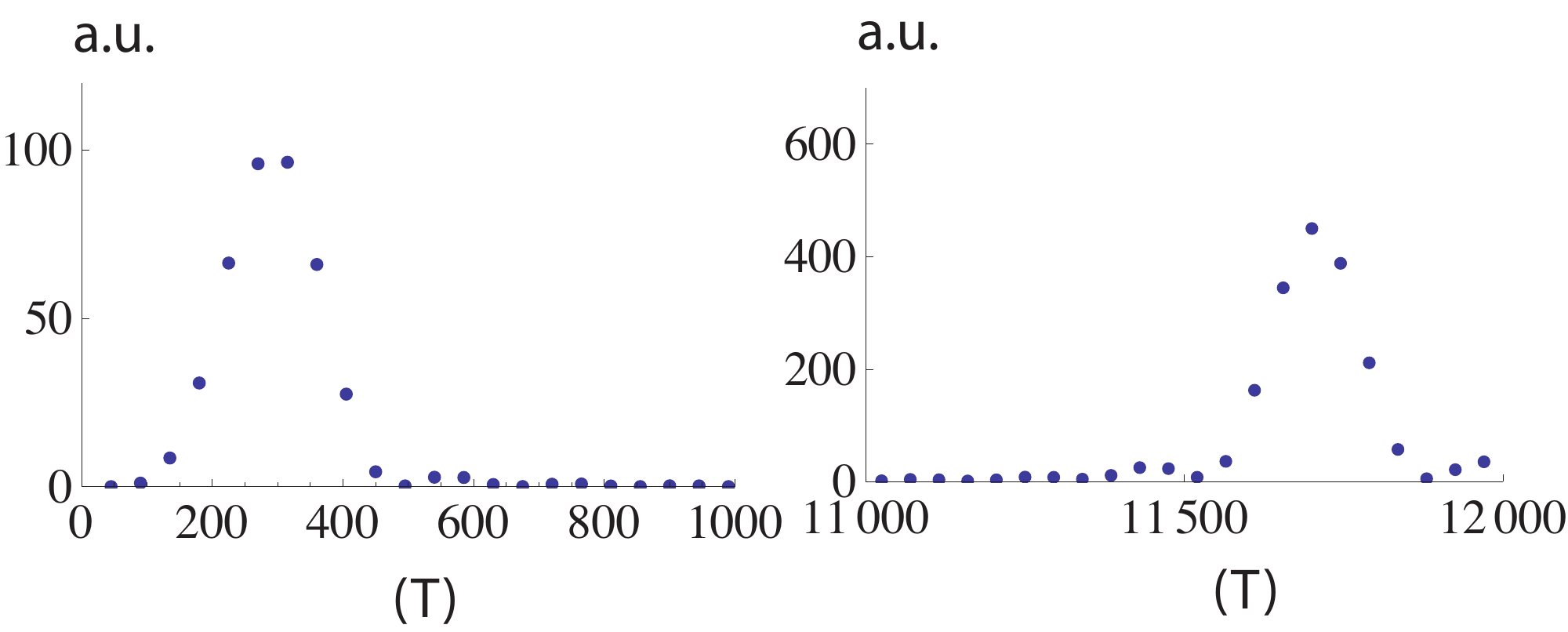}
\caption{(Color online) The same as in Fig.~\ref{fig:gap5-1} corresponding to 16\%  doping  but with a DDW gap of $ 15 meV$ and disorder $V_{0}=0.2 t$.}
\label{fig:gap15-1}
\end{center}
\end{figure}

\begin{figure}
\begin{center}
\includegraphics[width=\linewidth]{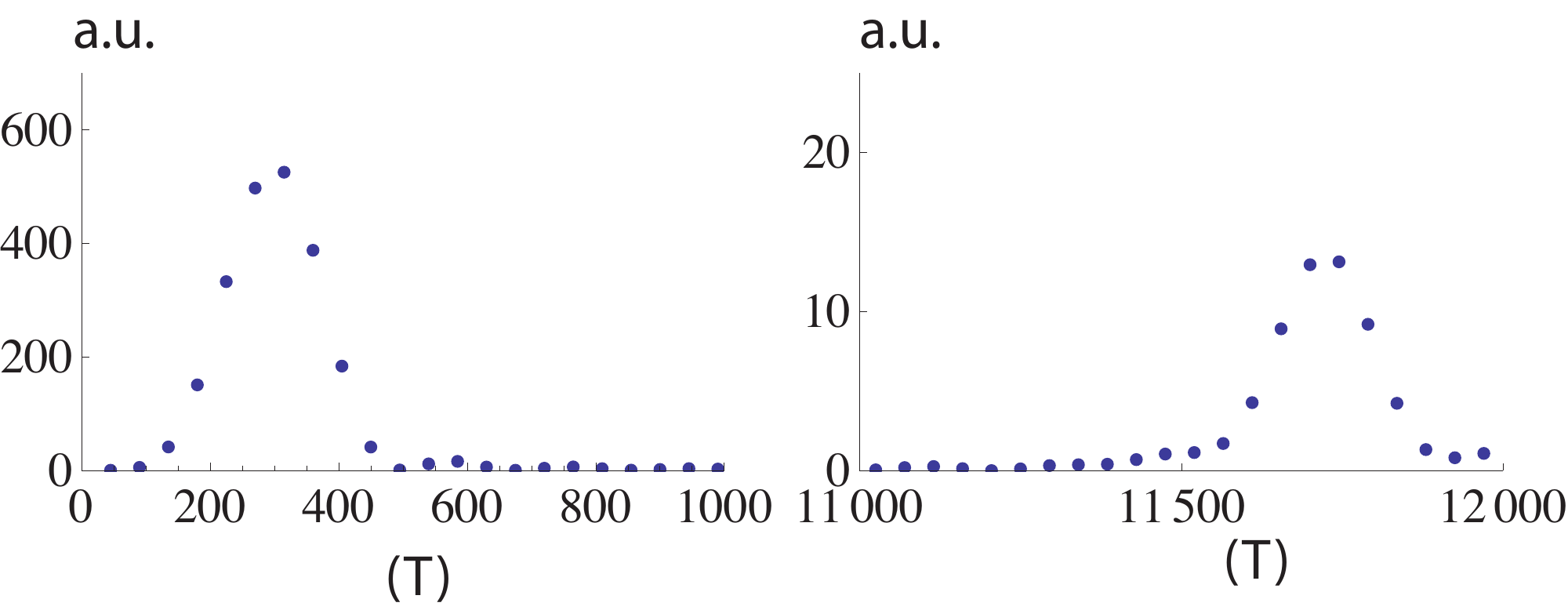}
\caption{(Color online) The same as in Fig.~\ref{fig:gap5-1} corresponding to 16\%  doping  but with a DDW gap of $ 15 meV$ and disorder $V_{0}=0.6 t$.}
\label{fig:gap15-3}
\end{center}
\end{figure}

\begin{figure}
\begin{center}
\includegraphics[width=\linewidth]{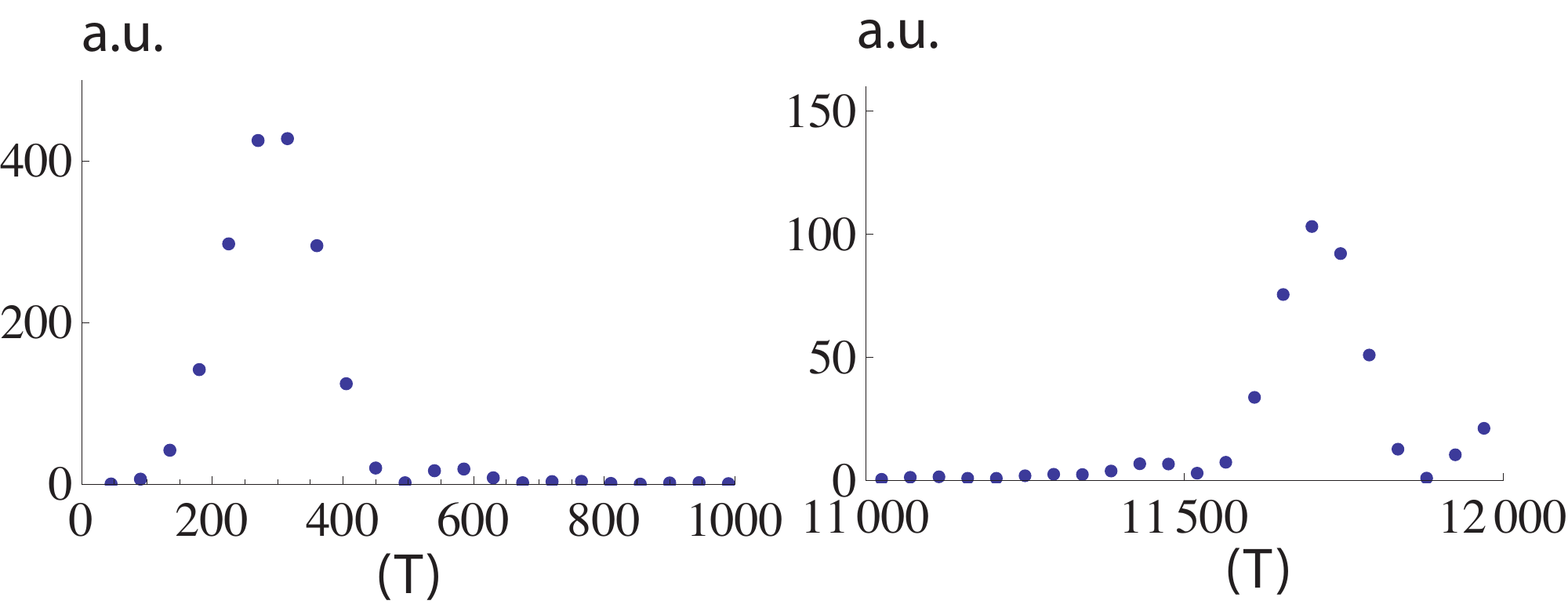}
\caption{(Color online) The same as in Fig.~\ref{fig:gap5-1} corresponding to 16\%  doping  but with a DDW gap of $ 30 meV$, and disorder $V_{0}=0.2 t$.}
\label{fig:gap30-1}
\end{center}
\end{figure}

\begin{figure}
\begin{center}
\includegraphics[width=\linewidth]{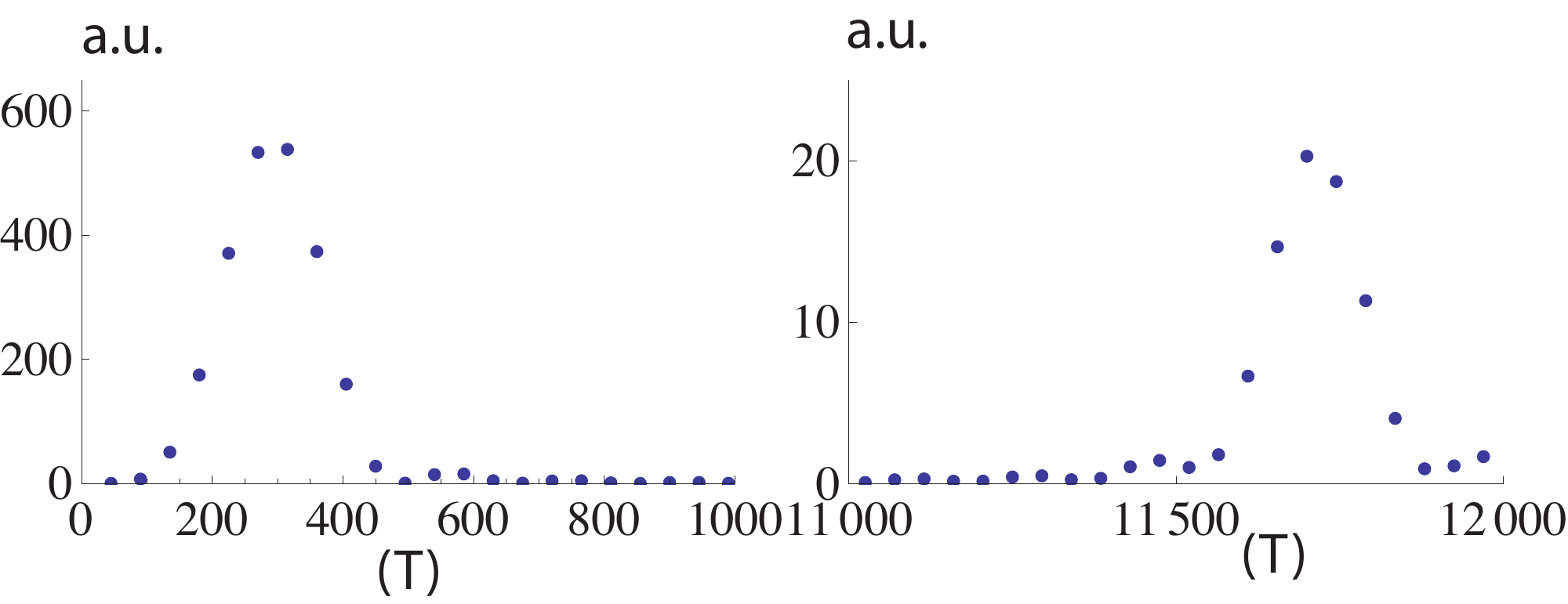}
\caption{(Color online) The same as in Fig.~\ref{fig:gap5-1} corresponding to 16\%  doping  but with a DDW gap of $ 30 meV$ and disorder $V_{0}=0.4 t$.}
\label{fig:gap30-2}
\end{center}
\end{figure}

\begin{figure}
\begin{center}
\includegraphics[width=\linewidth]{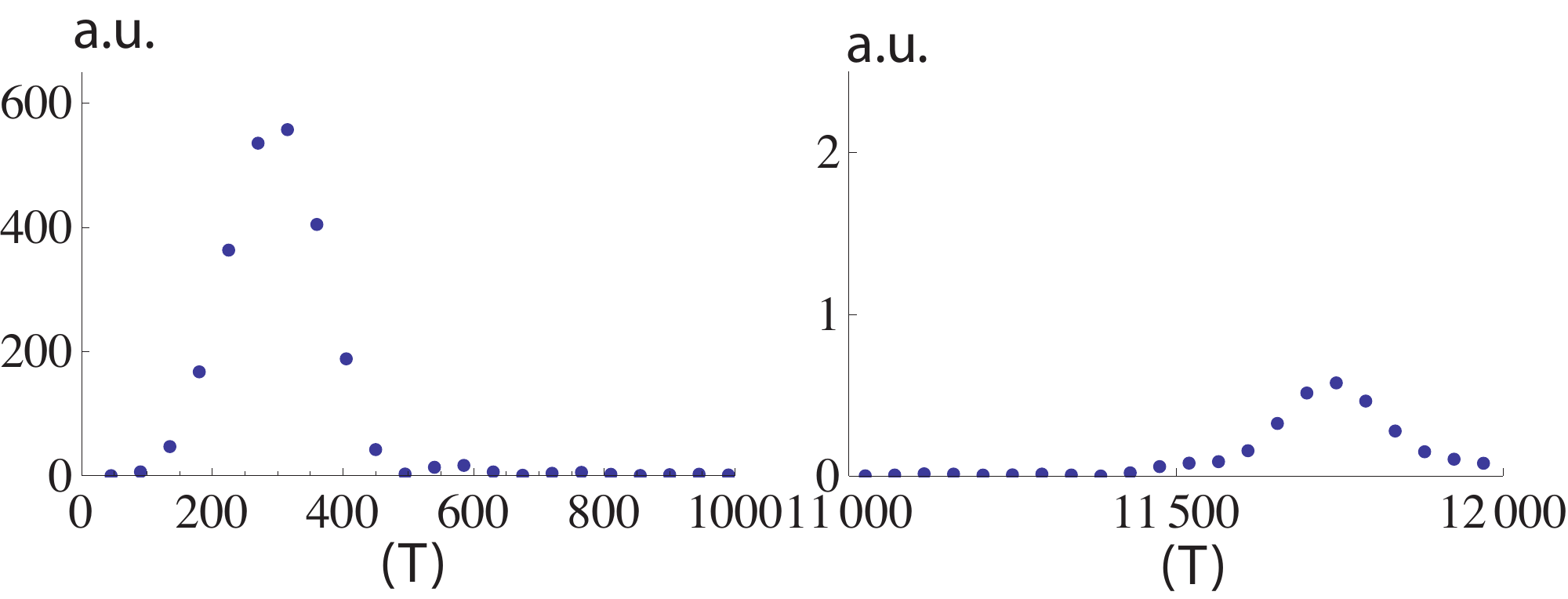}
\caption{(Color online) The same as in Fig.~\ref{fig:gap5-1} corresponding to 16\%  doping  but with a DDW gap of $ 30 meV$ and disorder $V_{0}=0.6 t$.}
\label{fig:gap30-3}
\end{center}
\end{figure}

In Figs.~\ref{fig:gap5-1} through~\ref{fig:gap5-3} the results for 17\% doping for a $5meV$ gap and varying degrees of disorder are shown.  Both the slow oscillation at a frequency $~290 T$ corresponding to the small hole pocket and $~11,700 T$ corresponding to the large hole pocket, as schematically sketched in Fig.~\ref{fig:Three-Pocket} in the extended Brillouin zone, can be seen. Note that partitioning of the spectral weight between the peaks changes as the degree of disorder is increased. If we change the value of the gap to $10 meV$, shown in Figs.~\ref{fig:gap10-1} through~\ref{fig:gap10-3}, the overall picture remains the same, although the slower frequency peak is a bit more dominant, as the magnetic breakdown is a little less probable. For 16\% doping similar calculation with gaps of $15 meV$ and $30 meV$ also show some evidence of magnetic breakdown depending on the disorder level, particularly seen in $15meV$ data in Figs.~\ref{fig:gap15-1} though~\ref{fig:gap15-3}. On the other hand, the evidence of magnetic breakdown is much weaker in the $30 meV$ data shown in Figs.~\ref{fig:gap30-1} through~\ref{fig:gap30-3}.

\begin{figure}
\begin{center}
\includegraphics[width=\linewidth]{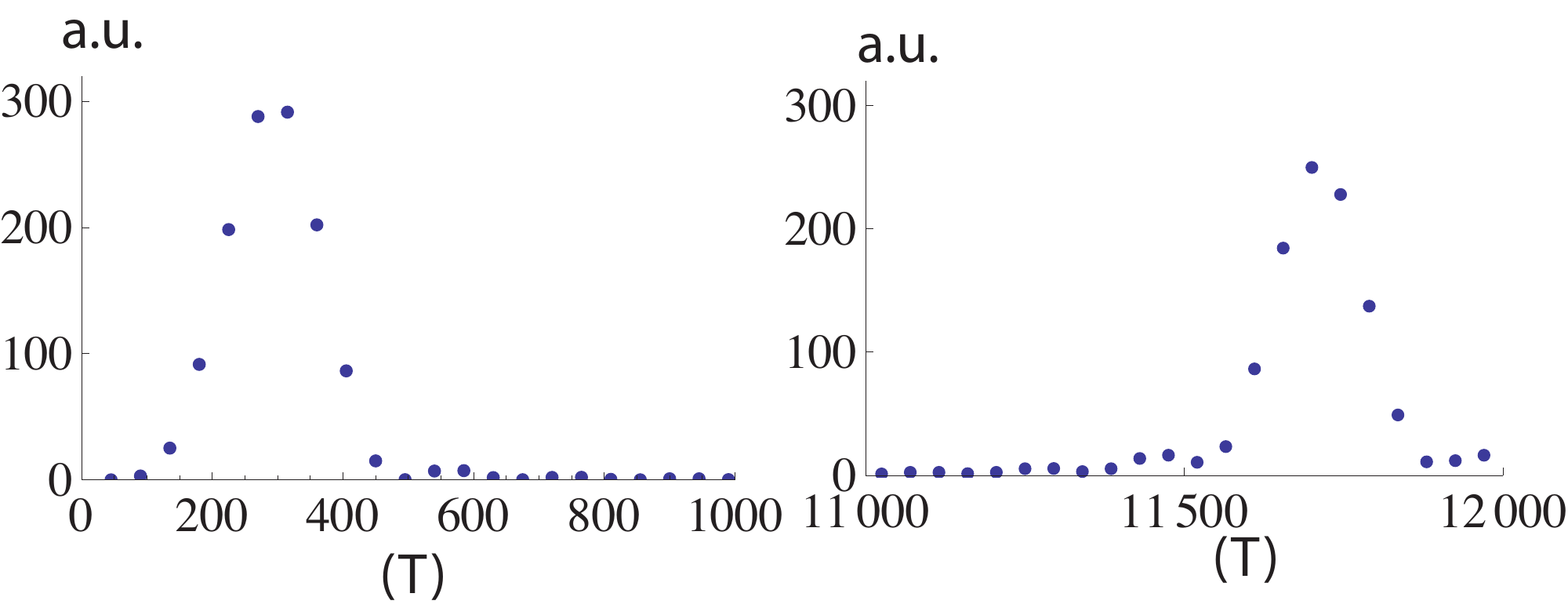}
\caption{(Color online) The same as in Fig.~\ref{fig:gap5-1} corresponding to 16\%  doping  but with a DDW gap of $ 15 meV$ and disorder $V_{0}=0.4 t$.}
\label{fig:gap15-2}
\end{center}
\end{figure}

It is important to note that in none of these calculations one finds any evidence of the electron pocket centered at $(\pi,0)$ and its symmetry counterparts, which should roughly correspond to a frequency of $2700 T$. This is in part due to the fact that the effect of disorder is stronger on  the electron pocket~\cite{Eun:2010}  and in part due to the fact that at the breakdown junctions transmission coefficient is larger than the reflection coefficient because it entails a large ($\pi/2$) change in the direction of the momentum; see Fig.~\ref{fig:Three-Pocket}.

\section{Oscillations in $c$-axis resistivity}
The Pichard-Landauer formula was calculated for conductance oscillations in the $ab$-plane, while the actual measurements in NCCO are carried for the $c$-axis resistivity. It is therefore necessary to relate the two to compare with experiments.
A simple description for a strongly layered material    can be obtained by modifying an argument of Kumar and Jayannavar.~\cite{Kumar:1992} An applied electric field, $E$, along the direction perpendicular to the planes will result in a chemical potential difference 
\begin{equation}
\Delta \mu = e d E,
\end{equation}
where $d$ is the distance between the two planes of an  unit cell. The corresponding current, $j_{c}$,
is ($\varepsilon_{F}$ is the Fermi energy)
\begin{equation}
j_{c} = e \left[\Delta \mu g_{2D}(\varepsilon_{F},H)\right] \gamma,
\end{equation}
since $\Delta \mu g_{2D}(\varepsilon_{F},H)$ is the number of unoccupied states to which an electron can scatter, while $\gamma$ is the scattering rate between the planes of a unit cell. Here, we have included a possible oscillatory dependence of the the two-dimensional density of states, $g_{2D}(\varepsilon_{F},H)$, that gives rise to Shubnikov-de Haas oscillations in the $ab$-plane. Thus,
\begin{equation}
\rho_{c}=\frac{E}{j_{c}}=\frac{1}{e^{2}dg_{2D}(\varepsilon_{F})\gamma}
\end{equation}
There is an implicit assumption: an electron from a given plane makes a transition to a continuum of available states with a finite density at the Fermi surface in the next plane. We are not interested in the Rabi oscillations  between two discrete states, a process that cannot lead to resistivity.

The measured $ab$-plane resistivity is of
the order $10\mu\Omega$-cm as compared $\Omega$-cm for the $c$-axis
resistivity even at optimum doping,~\cite{Armitage:2009} which
allows us to make an adiabatic approximation. Because an electron spends much of its time in the plane, making only infrequent hops between the planes, we can adiabatically decouple these two processes.  The slower motion along the $c$-axis can be formulated in terms of a $2\times 2$ matrix for each  parallel   wave vector ${\bf k}_{\parallel}$ after integrating out the  planar modes. For simplicity, we are assuming that the $c$-axis warping is negligible, so there are only two available states of the electron corresponding to its locations in the two planes.  The excitations  in a plane  close to the Fermi surface, $k_{\parallel}\approx k_{F,\parallel}$, can be approximated by a bosonic heat bath of particle-hole excitations. In this language, the problem maps on to a two-state  Hamiltonian 
\begin{equation}
H=-t_{c}\sigma_{x}+\sum_{j}\hbar\omega_{j}b^{\dagger}_{j}b_{j}+\frac{\sigma_{z}}{2}\sum_{j}f_{j}(b^{\dagger}_{j}+b_{j}),
\end{equation}
where $\sigma$'s are the standard Pauli matrices and $t_{c}$ is the hopping matrix element between the nearest neighbor planes..
Given the simplification, the sum over ${\bf k}_{\parallel}$ is superfluous, and the problem then maps on to a much studied model of a two-level system coupled to an Ohmic heat bath.~\cite{Chakravarty:1984,*Leggett:1987} The Ohmic nature follows from the fermionic nature of the bath.~\cite{Chang:1985}The effect of the bath on the transition between the planest is summarized by a spectral function,
\begin{equation}
J(\omega)=\frac{\pi}{2}\sum_{j} f_{j}^{2}\delta(\omega-\omega_{j}).
\end{equation}
For a fermionic bath, we can choose
\begin{equation}
J(\omega)=\begin{cases} 2\pi\alpha \omega, \; &\omega \ll \omega_{c}\\
                                              0, \; &\omega \gg \omega_{c}
                                              \end{cases}
\end{equation}
where $\omega_{c}$ is a high frequency cutoff, which is of the order of  $\omega_{c}= 2/\tau_{ab}$, where $\tau_{ab}$ is of the order of the planar relaxation time. For a Fermi bath, the parameter $\alpha$  is necessarily restricted to the range $0\le \alpha \le 1$.~\cite{Chang:1985} Moreover, for coherent oscillations we must have $\alpha < 1/2$.~\cite{Chakravarty:1984} However, we shall leave $\alpha$ as an adjustable parameter, presumably less than or equal to $1/2$ to be consistent with our initial assumptions. While a similar treatment is possible for a non-Fermi liquid,~\cite{Chakravarty:1994}
 the present discussion is entirely within the Fermi liquid theory.

The quantity $\gamma$ is the  interplanar tunneling tunneling rate renormalized by the particle-hole excitations close to the planar Fermi surface and can be easily seen to be~\cite{Chakravarty:1984}  
\begin{equation}
\gamma= \frac{2t_{c}}{\hbar}\left(\frac{2t_{c}}{\hbar\omega_{c}}\right)^{\frac{\alpha}{1-\alpha}}.
\end{equation}
The $c$-axis resistivity is then
\begin{equation}
\rho_{c}= \frac{\hbar}{e^{2}}\frac{1}{d g_{2D}(\varepsilon_{F},H)\hbar\omega_{c}}\left(\frac{\hbar\omega_{c}}{2t_{c}}\right)^{\frac{1}{1-\alpha}}
\end{equation}
This equation can be further simplified by expressing it as a ratio of $\rho_{c}/\rho_{ab}$, but this is unnecessary. Two important qualitative points are:  $\rho_{c}$ is far greater than $\rho_{ab}$ and the root of the  quantum oscillations  of $\rho_{c}$ is  quantum oscillations of the planar density of states.
\section{Conclusions}
We have shown that a qualitatively consistent physical picture for quantum oscillations can be provided with a simple set of assumptions
involving reconstruction of the Fermi surface due to density wave order. Although the specific order considered here was the DDW,
we have shown previously that at the mean field level a very similar picture can be provided by a two-fold commensurate spin density wave (SDW).~\cite{Eun:2010}
Thus, it appeared unnecessary to repeat the same calculations using the SDW order.

In YBCO, studies involving tilted field seems to rule out  triplet order parameter, hence SDW.~\cite{Ramshaw:2011,*Sebastian:2011} Moreover, from NMR measurements at high fields, there appears to be no evidence of a 
static spin density wave order in YBCO.~\cite{Julien:2011} Similarly there is no evidence of SDW order in fields as high as $23.2T$ in $\mathrm{YBa_{2}Cu_{4}O_{8}}$~\cite{Zheng:1999}, while quantum oscillations are clearly observed in this material.~\cite{yelland:2008,*Bangura:2008}  Also no such evidence of SDW  is found up to $44T$ in $\mathrm{Bi_{2}Sr_{2-x} La_{x}CuO_{6+\delta}}$.~\cite{Kawasaki:2011} At present, results from high field NMR  in NCCO does not exist, but measurements are  in progress.~\cite{Brown:2011} It is unlikely that such static SDW order will
be revealed in these measurements. This conjecture is based on the zero field neutron scattering measurements which indicate very small spin-spin correlation length in
the relevant doping regime.~\cite{Motoyama:2007} A long range SDW order cannot appear merely by applying high magnetic fields, which is energetically a weak perturbation even for $45 T$ field.~\cite{Nguyen:2002} 

As to singlet order, most likely relevant to the observation of quantum oscillations,~\cite{Garcia-Aldea:2011,*Norman:2011,*Ramazashvili:2011}  charge density wave  is a possibility, which has recently found some support in the 
high field NMR measurements  in YBCO.~\cite{Julien:2011} But since the mechanism involves oxygen chains, it is unlikely that the corresponding NMR measurements in NCCO will find such a charge order.
As to  singlet DDW, there are two neutron scattering measurements that seem to provide evidence for it.~\cite{Mook:2002,*Mook:2004} However, these measurements have not been confirmed by further independent experiments. However, DDW order should be considerably hidden in NMR involving nuclei at high symmetry points, because the orbital currents should cancel.

A mysterious feature of quantum oscillations in YBCO is the fact that only one type of Fermi pockets are observed. If two-fold commensurate density wave is 
the mechanism, this will violate the Luttinger sum rule.~\cite{Luttinger:1960,*Chubukov:1997,*Altshuler:1998,*Chakravarty:2008b} We have previously provided an explanation for this phenomenon in terms of disorder arising from both
defects and vortex scattering in the vortex liquid phase;~\cite{Jia:2009} however, the arguments are not unassailable. In contrast, for NCCO, the experimental results are 
quite consistent with the simple theory discussed above. We have not addressed AMRO in NCCO, as the data seem to be somewhat anomalous,~\cite{Kartsovnik:2011} although within the Fermi liquid framework
discussed here, it should be possible to address this effect in the future. 

The basic question as to  why Fermi liquid concepts should apply remains an important  unsolved mystery.~\cite{Chakravarty:2011} It is possible that if the state
revealed by applying a high magnetic field has a broken symmetry with an order parameter (hence a gap), the low energy excitations will be quasiparticle-like, not a spectra with a  
branch cut, as in variously proposed strange metal phases. In this respect, the notion of a hidden Fermi liquid may be relevant.~\cite{Casey:2011}

\section{Acknowledgments}
We thank Mark Kartsovnik, Stuart Brown, Marc-Henri Julien, Brad Ramshaw, and Cyril Proust for keeping us updated regarding their
latest experiments. In particular, we thank Marc-Henri Julien for sharing with us his unpublished  high field NMR results in YBCO.
This work is supported by NSF under the Grant DMR-1004520.
\appendix*
\section{The derivation of the transfer matrix}
The DDW Hamiltonian in real space is
\begin{widetext}
\begin{equation}
\begin{split}
    H=\sum_{\vec{i}}\epsilon_\vec{i}c_\vec{i}^\dag c_\vec{i}
-t    \sum_{\langle\vec{i},\vec{j}\rangle} ~\mathrm{e}^{ia_{\vec{i},\vec{j}}}c_\vec{i}^\dag c_\vec{j} 
-t'   \sum_{\langle\vec{i},\vec{j}\rangle'} ~\mathrm{e}^{ia_{\vec{i},\vec{j}}}c_\vec{i}^\dag c_\vec{j} 
-t''  \sum_{\langle\vec{i},\vec{j}\rangle''} ~\mathrm{e}^{ia_{\vec{i},\vec{j}}}c_\vec{i}^\dag c_\vec{j}\\
+	\sum_{\vec{i}} ~ \frac{i W_0}{4}(-1)^{n+m} c_\vec{i}^\dag c_\vec{\vec{i}+\hat{\vec{x}}}
-\sum_{\vec{i}}  ~ \frac{i W_0}{4}(-1)^{n+m} c_\vec{i}^\dag c_\vec{\vec{i}+\hat{\vec{y}}}
	+ h.c.
\end{split}
\end{equation}
\end{widetext}
Here, $~\mathrm{e}^{ia_{\vec{i},\vec{j}}}$ is the  Peierls phase due to the magnetic field. The summation notations are as follows: $\langle\vec{i},\vec{j}\rangle$, $\langle\vec{i},\vec{j}\rangle'$, and $\langle\vec{i},\vec{j}\rangle''$ imply sum over  nearest-neighbor, next-nearest-neighbor, and the third-nearest-neighobor sites, respectively. For example, with the lattice constant set to unity, $\langle\vec{i},\vec{j}\rangle$ is satisfied when $\vec{i}=\vec{j} \pm \hat{\vec{x}}$ or $\vec{i}=\vec{j} \pm \hat{\vec{y}}$. Likewise, $\langle\vec{i},\vec{j}\rangle'$ requires $\vec{i}=\vec{j} + \hat{\vec{x}} \pm \hat{\vec{y}}$ or $\vec{i}=\vec{j} - \hat{\vec{x}} \pm \hat{\vec{y}}$ and $\langle\vec{i},\vec{j}\rangle''$ requires $\vec{i}=\vec{j} \pm 2\hat{\vec{x}}$ or $\vec{i}=\vec{j} \pm 2\hat{\vec{y}}$.
Here $W_{0}$ is the DDW gap and $\vec{i}=(n,m)$. Consider an eigen state $|\Psi\rangle$ with an energy eigenvalue $E$: $H|\Psi\rangle=E|\Psi\rangle$, where $|\Psi\rangle=\sum_{\vec{i}}\psi(\vec{i})|\vec{i}\rangle$; the amplitude at a site is $\psi(\vec{i})$. Then the Schr\"odinger equation can be written in terms of the amplitudes $\psi_{n}(m)$ of the $n$-th slice for all values of $m=1,2,\ldots M$:
\begin{widetext}
\begin{equation}
\begin{split}
E \psi_{n}(m) = \epsilon_\vec{i} \psi_{n}(m) 
-t\left[\psi_{n+1}(m)+\psi_{n-1}(m)+\me^{-\mi n\phi} \psi_{n}(m+1)+\me^{\mi n\phi} \psi_{n}(m-1) \right] \\
-t'\left[\me^{\mi(-n-\frac{1}{2})\phi}  \psi_{n+1}(m+1)
+ \me^{\mi(n+\frac{1}{2})\phi} \psi_{n-1}(m+1)
+ \me^{\mi(n+\frac{1}{2})\phi} \psi_{n+1}(m-1)
+ \me^{\mi(-n-\frac{1}{2})\phi} \psi_{n-1}(m-1) \right] \\
-t''\left[ \psi_{n+2}(m)+\psi_{n-2}(m)+ 
\me^{-\mi 2 n\phi} \psi_{n}(m+2)+ \me^{\mi 2 n\phi} \psi_{n}(m-2) \right] \\
+\frac{i W_0}{4}(-1)^{n+m} \left[\psi_{n+1}(m)+\psi_{n-1}(m)\right] 
-\frac{i W_0}{4}(-1)^{n+m} \left[\me^{-\mi n\phi} \psi_{n}(m+1)+\me^{\mi n\phi} \psi_{n}(m-1)\right]
\end{split}
\end{equation}
\end{widetext}
With periodic boundary condition along the y-axis i.e. $\psi_n (M+1)=\psi_n (1)$, the   Schr\"odinger equation can be expressed as a matrix equation:
\begin{equation}
0 = -U_n \psi_{n+2} + A_n \psi_{n+1} + 
B_n \psi_{n} + C_n \psi_{n-1} + D_n \psi_{n-2},
\end{equation}
where $U_{n}$, $A_{n}$, $B_{n}$, $C_{n}$, and $D_{n}$ are $M\times M$ matrices defined in
equations following Eq.~\ref{Eq:transfermatrix}. Now we can solve the Schrodinger equation for $\psi_{n+2}$ to obtain $\psi_{n+2} = U_n^{-1}(A_n \psi_{n+1} + B_n \psi_{n} + C_n \psi_{n-1} + D_n \psi_{n-2})$. Then the amplitudes at a  set of four successive slices,  $\psi_{n-1}$ through $\psi_{n+2}$, can be written in terms of the amplitudes of a previous set of four successive slices,  $\psi_{n-2}$ through $\psi_{n+1}$. Thus, the transfer matrix in the main text follows.


\begin{thebibliography}{43}%
\makeatletter
\providecommand \@ifxundefined [1]{%
 \@ifx{#1\undefined}
}%
\providecommand \@ifnum [1]{%
 \ifnum #1\expandafter \@firstoftwo
 \else \expandafter \@secondoftwo
 \fi
}%
\providecommand \@ifx [1]{%
 \ifx #1\expandafter \@firstoftwo
 \else \expandafter \@secondoftwo
 \fi
}%
\providecommand \natexlab [1]{#1}%
\providecommand \enquote  [1]{``#1''}%
\providecommand \bibnamefont  [1]{#1}%
\providecommand \bibfnamefont [1]{#1}%
\providecommand \citenamefont [1]{#1}%
\providecommand \href@noop [0]{\@secondoftwo}%
\providecommand \href [0]{\begingroup \@sanitize@url \@href}%
\providecommand \@href[1]{\@@startlink{#1}\@@href}%
\providecommand \@@href[1]{\endgroup#1\@@endlink}%
\providecommand \@sanitize@url [0]{\catcode `\\12\catcode `\$12\catcode
  `\&12\catcode `\#12\catcode `\^12\catcode `\_12\catcode `\%12\relax}%
\providecommand \@@startlink[1]{}%
\providecommand \@@endlink[0]{}%
\providecommand \url  [0]{\begingroup\@sanitize@url \@url }%
\providecommand \@url [1]{\endgroup\@href {#1}{\urlprefix }}%
\providecommand \urlprefix  [0]{URL }%
\providecommand \Eprint [0]{\href }%
\@ifxundefined \urlstyle {%
  \providecommand \doi  [0]{\begingroup \@sanitize@url \@doi}%
  \providecommand \@doi [1]{\endgroup \@@startlink {\doibase
  #1}doi:\discretionary {}{}{}#1\@@endlink }%
}{%
  \providecommand \doi  [0]{doi:\discretionary{}{}{}\begingroup
  \urlstyle{rm}\Url }%
}%
\providecommand \doibase [0]{http://dx.doi.org/}%
\providecommand \Doi [0]{\begingroup \@sanitize@url \@Doi }%
\providecommand \@Doi  [1]{\endgroup\@@startlink{\doibase#1}\@@Doi}%
\providecommand \@@Doi [1]{#1\@@endlink}%
\providecommand \selectlanguage [0]{\@gobble}%
\providecommand \bibinfo  [0]{\@secondoftwo}%
\providecommand \bibfield  [0]{\@secondoftwo}%
\providecommand \translation [1]{[#1]}%
\providecommand \BibitemOpen [0]{}%
\providecommand \bibitemStop [0]{}%
\providecommand \bibitemNoStop [0]{.\EOS\space}%
\providecommand \EOS [0]{\spacefactor3000\relax}%
\providecommand \BibitemShut  [1]{\csname bibitem#1\endcsname}%
\bibitem [{\citenamefont {Doiron-Leyraud}\ \emph {et~al.}(2007)\citenamefont
  {Doiron-Leyraud}, \citenamefont {Proust}, \citenamefont {LeBoeuf},
  \citenamefont {Levallois}, \citenamefont {Bonnemaison}, \citenamefont
  {Liang}, \citenamefont {Bonn}, \citenamefont {Hardy},\ and\ \citenamefont
  {Taillefer}}]{Doiron-Leyraud:2007}%
  \BibitemOpen
  \bibfield  {author} {\bibinfo {author} {\bibfnamefont {N.}~\bibnamefont
  {Doiron-Leyraud}}, \bibinfo {author} {\bibfnamefont {C.}~\bibnamefont
  {Proust}}, \bibinfo {author} {\bibfnamefont {D.}~\bibnamefont {LeBoeuf}},
  \bibinfo {author} {\bibfnamefont {J.}~\bibnamefont {Levallois}}, \bibinfo
  {author} {\bibfnamefont {J.-B.}\ \bibnamefont {Bonnemaison}}, \bibinfo
  {author} {\bibfnamefont {R.}~\bibnamefont {Liang}}, \bibinfo {author}
  {\bibfnamefont {D.~A.}\ \bibnamefont {Bonn}}, \bibinfo {author}
  {\bibfnamefont {W.~N.}\ \bibnamefont {Hardy}}, \ and\ \bibinfo {author}
  {\bibfnamefont {L.}~\bibnamefont {Taillefer}},\ }\href@noop {} {\bibfield
  {journal} {\bibinfo  {journal} {Nature},\ }\textbf {\bibinfo {volume}
  {447}},\ \bibinfo {pages} {565} (\bibinfo {year} {2007})}\BibitemShut
  {NoStop}%
\bibitem [{\citenamefont {Riggs}\ \emph {et~al.}(2011)\citenamefont {Riggs},
  \citenamefont {Vafek}, \citenamefont {Kemper}, \citenamefont {Betts},
  \citenamefont {Migliori}, \citenamefont {Hardy}, \citenamefont {Liang},
  \citenamefont {Bonn},\ and\ \citenamefont {Boebinger}}]{Riggs:2011}%
  \BibitemOpen
  \bibfield  {author} {\bibinfo {author} {\bibfnamefont {S.~C.}\ \bibnamefont
  {Riggs}}, \bibinfo {author} {\bibfnamefont {O.}~\bibnamefont {Vafek}},
  \bibinfo {author} {\bibfnamefont {J.~B.}\ \bibnamefont {Kemper}}, \bibinfo
  {author} {\bibfnamefont {J.}~\bibnamefont {Betts}}, \bibinfo {author}
  {\bibfnamefont {A.}~\bibnamefont {Migliori}}, \bibinfo {author}
  {\bibfnamefont {W.~N.}\ \bibnamefont {Hardy}}, \bibinfo {author}
  {\bibfnamefont {R.}~\bibnamefont {Liang}}, \bibinfo {author} {\bibfnamefont
  {D.~A.}\ \bibnamefont {Bonn}}, \ and\ \bibinfo {author} {\bibfnamefont
  {G.}~\bibnamefont {Boebinger}},\ }\href@noop {} {\bibfield  {journal}
  {\bibinfo  {journal} {Nat. Phys.},\ }\textbf {\bibinfo {volume} {7}},\
  \bibinfo {pages} {332} (\bibinfo {year} {2011})}\BibitemShut {NoStop}%
\bibitem [{\citenamefont {Chakravarty}(2008)}]{Chakravarty:2008}%
  \BibitemOpen
  \bibfield  {author} {\bibinfo {author} {\bibfnamefont {S.}~\bibnamefont
  {Chakravarty}},\ }\href@noop {} {\bibfield  {journal} {\bibinfo  {journal}
  {Science},\ }\textbf {\bibinfo {volume} {319}},\ \bibinfo {pages} {735}
  (\bibinfo {year} {2008})}\BibitemShut {NoStop}%
\bibitem [{\citenamefont {Chakravarty}\ and\ \citenamefont
  {Kee}(2008)}]{Chakravarty:2008b}%
  \BibitemOpen
  \bibfield  {author} {\bibinfo {author} {\bibfnamefont {S.}~\bibnamefont
  {Chakravarty}}\ and\ \bibinfo {author} {\bibfnamefont {H.-Y.}\ \bibnamefont
  {Kee}},\ }\href@noop {} {\bibfield  {journal} {\bibinfo  {journal} {Proc.
  Natl. Acad. Sci. USA},\ }\textbf {\bibinfo {volume} {105}},\ \bibinfo {pages}
  {8835} (\bibinfo {year} {2008})}\BibitemShut {NoStop}%
\bibitem [{\citenamefont {Dimov}\ \emph {et~al.}(2008)\citenamefont {Dimov},
  \citenamefont {Goswami}, \citenamefont {Jia},\ and\ \citenamefont
  {Chakravarty}}]{Dimov:2008}%
  \BibitemOpen
  \bibfield  {author} {\bibinfo {author} {\bibfnamefont {I.}~\bibnamefont
  {Dimov}}, \bibinfo {author} {\bibfnamefont {P.}~\bibnamefont {Goswami}},
  \bibinfo {author} {\bibfnamefont {X.}~\bibnamefont {Jia}}, \ and\ \bibinfo
  {author} {\bibfnamefont {S.}~\bibnamefont {Chakravarty}},\ }\href@noop {}
  {\bibfield  {journal} {\bibinfo  {journal} {Phys. Rev. B},\ }\textbf
  {\bibinfo {volume} {78}},\ \bibinfo {pages} {134529} (\bibinfo {year}
  {2008})}\BibitemShut {NoStop}%
\bibitem [{\citenamefont {Millis}\ and\ \citenamefont
  {Norman}(2007)}]{Millis:2007}%
  \BibitemOpen
  \bibfield  {author} {\bibinfo {author} {\bibfnamefont {A.~J.}\ \bibnamefont
  {Millis}}\ and\ \bibinfo {author} {\bibfnamefont {M.~R.}\ \bibnamefont
  {Norman}},\ }\href@noop {} {\bibfield  {journal} {\bibinfo  {journal} {Phys.
  Rev. B},\ }\textbf {\bibinfo {volume} {76}},\ \bibinfo {pages} {220503}
  (\bibinfo {year} {2007})}\BibitemShut {NoStop}%
\bibitem [{\citenamefont {Yao}\ \emph {et~al.}(2011)\citenamefont {Yao},
  \citenamefont {Lee},\ and\ \citenamefont {Kivelson}}]{Yao:2011}%
  \BibitemOpen
  \bibfield  {author} {\bibinfo {author} {\bibfnamefont {H.}~\bibnamefont
  {Yao}}, \bibinfo {author} {\bibfnamefont {D.-H.}\ \bibnamefont {Lee}}, \ and\
  \bibinfo {author} {\bibfnamefont {S.~A.}\ \bibnamefont {Kivelson}},\
  }\href@noop {} {\enquote {\bibinfo {title} {{Fermi-surface reconstruction in
  a smectic phase of a high temperature superconductor}},}\ } (\bibinfo {year}
  {2011}),\ \Eprint {http://arxiv.org/abs/1103.211v1 [cond-mat]}
  {arXiv:1103.211v1 [cond-mat]} \BibitemShut {NoStop}%
\bibitem [{\citenamefont {Helm}\ \emph {et~al.}(2009)\citenamefont {Helm},
  \citenamefont {Kartsovnik}, \citenamefont {Bartkowiak}, \citenamefont
  {Bittner}, \citenamefont {Lambacher}, \citenamefont {Erb}, \citenamefont
  {Wosnitza},\ and\ \citenamefont {Gross}}]{Helm:2009}%
  \BibitemOpen
  \bibfield  {author} {\bibinfo {author} {\bibfnamefont {T.}~\bibnamefont
  {Helm}}, \bibinfo {author} {\bibfnamefont {M.~V.}\ \bibnamefont
  {Kartsovnik}}, \bibinfo {author} {\bibfnamefont {M.}~\bibnamefont
  {Bartkowiak}}, \bibinfo {author} {\bibfnamefont {N.}~\bibnamefont {Bittner}},
  \bibinfo {author} {\bibfnamefont {M.}~\bibnamefont {Lambacher}}, \bibinfo
  {author} {\bibfnamefont {A.}~\bibnamefont {Erb}}, \bibinfo {author}
  {\bibfnamefont {J.}~\bibnamefont {Wosnitza}}, \ and\ \bibinfo {author}
  {\bibfnamefont {R.}~\bibnamefont {Gross}},\ }\href@noop {} {\bibfield
  {journal} {\bibinfo  {journal} {Phys. Rev. Lett.},\ }\textbf {\bibinfo
  {volume} {103}},\ \bibinfo {pages} {157002} (\bibinfo {year}
  {2009})}\BibitemShut {NoStop}%
\bibitem [{\citenamefont {Armitage}\ \emph {et~al.}(2009)\citenamefont
  {Armitage}, \citenamefont {Fournier},\ and\ \citenamefont
  {Green}}]{Armitage:2009}%
  \BibitemOpen
  \bibfield  {author} {\bibinfo {author} {\bibfnamefont {N.~P.}\ \bibnamefont
  {Armitage}}, \bibinfo {author} {\bibfnamefont {P.}~\bibnamefont {Fournier}},
  \ and\ \bibinfo {author} {\bibfnamefont {R.~L.}\ \bibnamefont {Green}},\
  }\href@noop {} {\bibfield  {journal} {\bibinfo  {journal} {Rev. Mod. Phys.},\
  }\textbf {\bibinfo {volume} {82}},\ \bibinfo {pages} {2421} (\bibinfo {year}
  {2009})}\BibitemShut {NoStop}%
\bibitem [{\citenamefont {Eun}\ \emph {et~al.}(2010)\citenamefont {Eun},
  \citenamefont {Jia},\ and\ \citenamefont {Chakravarty}}]{Eun:2010}%
  \BibitemOpen
  \bibfield  {author} {\bibinfo {author} {\bibfnamefont {J.}~\bibnamefont
  {Eun}}, \bibinfo {author} {\bibfnamefont {X.}~\bibnamefont {Jia}}, \ and\
  \bibinfo {author} {\bibfnamefont {S.}~\bibnamefont {Chakravarty}},\
  }\href@noop {} {\bibfield  {journal} {\bibinfo  {journal} {Phys. Rev. B},\
  }\textbf {\bibinfo {volume} {82}},\ \bibinfo {pages} {094515} (\bibinfo
  {year} {2010})}\BibitemShut {NoStop}%
\bibitem [{\citenamefont {Helm}\ \emph {et~al.}(2010)\citenamefont {Helm},
  \citenamefont {Kartsovnik}, \citenamefont {Sheikin}, \citenamefont
  {Bartkowiak}, \citenamefont {Wolff-Fabris}, \citenamefont {Bittner},
  \citenamefont {Biberacher}, \citenamefont {Lambacher}, \citenamefont {Erb},
  \citenamefont {Wosnitza},\ and\ \citenamefont {Gross}}]{Helm:2010}%
  \BibitemOpen
  \bibfield  {author} {\bibinfo {author} {\bibfnamefont {T.}~\bibnamefont
  {Helm}}, \bibinfo {author} {\bibfnamefont {M.~V.}\ \bibnamefont
  {Kartsovnik}}, \bibinfo {author} {\bibfnamefont {I.}~\bibnamefont {Sheikin}},
  \bibinfo {author} {\bibfnamefont {M.}~\bibnamefont {Bartkowiak}}, \bibinfo
  {author} {\bibfnamefont {F.}~\bibnamefont {Wolff-Fabris}}, \bibinfo {author}
  {\bibfnamefont {N.}~\bibnamefont {Bittner}}, \bibinfo {author} {\bibfnamefont
  {W.}~\bibnamefont {Biberacher}}, \bibinfo {author} {\bibfnamefont
  {M.}~\bibnamefont {Lambacher}}, \bibinfo {author} {\bibfnamefont
  {A.}~\bibnamefont {Erb}}, \bibinfo {author} {\bibfnamefont {J.}~\bibnamefont
  {Wosnitza}}, \ and\ \bibinfo {author} {\bibfnamefont {R.}~\bibnamefont
  {Gross}},\ }\href@noop {} {\bibfield  {journal} {\bibinfo  {journal} {Phys.
  Rev. Lett.},\ }\textbf {\bibinfo {volume} {105}},\ \bibinfo {pages} {247002}
  (\bibinfo {year} {2010})}\BibitemShut {NoStop}%
\bibitem [{\citenamefont {Kartsovnik}\ and\ \citenamefont
  {et~al.}(2011)}]{Kartsovnik:2011}%
  \BibitemOpen
  \bibfield  {author} {\bibinfo {author} {\bibfnamefont {M.~V.}\ \bibnamefont
  {Kartsovnik}}\ and\ \bibinfo {author} {\bibnamefont {et~al.}},\ }\href@noop
  {} {\bibfield  {journal} {\bibinfo  {journal} {New J. Phys.},\ }\textbf
  {\bibinfo {volume} {13}},\ \bibinfo {pages} {015001} (\bibinfo {year}
  {2011})}\BibitemShut {NoStop}%
\bibitem [{\citenamefont {Chakravarty}\ \emph {et~al.}(2001)\citenamefont
  {Chakravarty}, \citenamefont {Laughlin}, \citenamefont {Morr},\ and\
  \citenamefont {Nayak}}]{Chakravarty:2001}%
  \BibitemOpen
  \bibfield  {author} {\bibinfo {author} {\bibfnamefont {S.}~\bibnamefont
  {Chakravarty}}, \bibinfo {author} {\bibfnamefont {R.~B.}\ \bibnamefont
  {Laughlin}}, \bibinfo {author} {\bibfnamefont {D.~K.}\ \bibnamefont {Morr}},
  \ and\ \bibinfo {author} {\bibfnamefont {C.}~\bibnamefont {Nayak}},\
  }\href@noop {} {\bibfield  {journal} {\bibinfo  {journal} {Phys. Rev. B},\
  }\textbf {\bibinfo {volume} {63}},\ \bibinfo {pages} {094503} (\bibinfo
  {year} {2001})}\BibitemShut {NoStop}%
\bibitem [{\citenamefont {Pavarini}\ \emph {et~al.}(2001)\citenamefont
  {Pavarini}, \citenamefont {Dasgupta}, \citenamefont {Saha-Dasgupta},
  \citenamefont {Jepsen},\ and\ \citenamefont {Andersen}}]{Pavarini:2001}%
  \BibitemOpen
  \bibfield  {author} {\bibinfo {author} {\bibfnamefont {E.}~\bibnamefont
  {Pavarini}}, \bibinfo {author} {\bibfnamefont {I.}~\bibnamefont {Dasgupta}},
  \bibinfo {author} {\bibfnamefont {T.}~\bibnamefont {Saha-Dasgupta}}, \bibinfo
  {author} {\bibfnamefont {O.}~\bibnamefont {Jepsen}}, \ and\ \bibinfo {author}
  {\bibfnamefont {O.~K.}\ \bibnamefont {Andersen}},\ }\href@noop {} {\bibfield
  {journal} {\bibinfo  {journal} {Phys. Rev. Lett.},\ }\textbf {\bibinfo
  {volume} {87}},\ \bibinfo {pages} {047003} (\bibinfo {year}
  {2001})}\BibitemShut {NoStop}%
\bibitem [{\citenamefont {Jia}\ \emph {et~al.}(2009)\citenamefont {Jia},
  \citenamefont {Goswami},\ and\ \citenamefont {Chakravarty}}]{Jia:2009}%
  \BibitemOpen
  \bibfield  {author} {\bibinfo {author} {\bibfnamefont {X.}~\bibnamefont
  {Jia}}, \bibinfo {author} {\bibfnamefont {P.}~\bibnamefont {Goswami}}, \ and\
  \bibinfo {author} {\bibfnamefont {S.}~\bibnamefont {Chakravarty}},\
  }\href@noop {} {\bibfield  {journal} {\bibinfo  {journal} {Phys. Rev. B},\
  }\textbf {\bibinfo {volume} {80}},\ \bibinfo {pages} {134503} (\bibinfo
  {year} {2009})}\BibitemShut {NoStop}%
\bibitem [{\citenamefont {Pichard}\ and\ \citenamefont
  {Andr\'e}(1986)}]{Pichard:1986}%
  \BibitemOpen
  \bibfield  {author} {\bibinfo {author} {\bibfnamefont {J.~L.}\ \bibnamefont
  {Pichard}}\ and\ \bibinfo {author} {\bibfnamefont {G.}~\bibnamefont
  {Andr\'e}},\ }\href@noop {} {\bibfield  {journal} {\bibinfo  {journal}
  {Europhys. Lett.},\ }\textbf {\bibinfo {volume} {2}},\ \bibinfo {pages} {477}
  (\bibinfo {year} {1986})}\BibitemShut {NoStop}%
\bibitem [{\citenamefont {Fisher}\ and\ \citenamefont
  {Lee}(1981)}]{Fisher:1981}%
  \BibitemOpen
  \bibfield  {author} {\bibinfo {author} {\bibfnamefont {D.~S.}\ \bibnamefont
  {Fisher}}\ and\ \bibinfo {author} {\bibfnamefont {P.~A.}\ \bibnamefont
  {Lee}},\ }\href@noop {} {\bibfield  {journal} {\bibinfo  {journal} {Phys.
  Rev. B},\ }\textbf {\bibinfo {volume} {23}},\ \bibinfo {pages} {6851}
  (\bibinfo {year} {1981})}\BibitemShut {NoStop}%
\bibitem [{\citenamefont {Kramer}\ and\ \citenamefont
  {Schreiber}(1996)}]{Kramer:1996}%
  \BibitemOpen
  \bibfield  {author} {\bibinfo {author} {\bibfnamefont {B.}~\bibnamefont
  {Kramer}}\ and\ \bibinfo {author} {\bibfnamefont {M.}~\bibnamefont
  {Schreiber}},\ }in\ \href@noop {} {\emph {\bibinfo {booktitle} {Computational
  Physics}}},\ \bibinfo {editor} {edited by\ \bibinfo {editor} {\bibfnamefont
  {K.~H.}\ \bibnamefont {Hoffmann}}\ and\ \bibinfo {editor} {\bibfnamefont
  {M.}~\bibnamefont {Schreiber}}}\ (\bibinfo  {publisher} {Springer},\ \bibinfo
  {address} {Berlin},\ \bibinfo {year} {1996})\ p.\ \bibinfo {pages}
  {166}\BibitemShut {NoStop}%
\bibitem [{\citenamefont {Kumar}\ and\ \citenamefont
  {Jayannavar}(1992)}]{Kumar:1992}%
  \BibitemOpen
  \bibfield  {author} {\bibinfo {author} {\bibfnamefont {N.}~\bibnamefont
  {Kumar}}\ and\ \bibinfo {author} {\bibfnamefont {A.~M.}\ \bibnamefont
  {Jayannavar}},\ }\href@noop {} {\bibfield  {journal} {\bibinfo  {journal}
  {Phys. Rev. B},\ }\textbf {\bibinfo {volume} {45}},\ \bibinfo {pages} {5001}
  (\bibinfo {year} {1992})}\BibitemShut {NoStop}%
\bibitem [{\citenamefont {Chakravarty}\ and\ \citenamefont
  {Leggett}(1984)}]{Chakravarty:1984}%
  \BibitemOpen
  \bibfield  {author} {\bibinfo {author} {\bibfnamefont {S.}~\bibnamefont
  {Chakravarty}}\ and\ \bibinfo {author} {\bibfnamefont {A.~J.}\ \bibnamefont
  {Leggett}},\ }\href@noop {} {\bibfield  {journal} {\bibinfo  {journal}
  {Physi. Rev. Lett.},\ }\textbf {\bibinfo {volume} {52}},\ \bibinfo {pages}
  {5} (\bibinfo {year} {1984})}\BibitemShut {NoStop}%
\bibitem [{\citenamefont {Leggett}\ \emph {et~al.}(1987)\citenamefont
  {Leggett}, \citenamefont {Chakravarty}, \citenamefont {Dorsey}, \citenamefont
  {Fisher}, \citenamefont {Garg},\ and\ \citenamefont
  {Zwerger}}]{Leggett:1987}%
  \BibitemOpen
  \bibfield  {author} {\bibinfo {author} {\bibfnamefont {A.~J.}\ \bibnamefont
  {Leggett}}, \bibinfo {author} {\bibfnamefont {S.}~\bibnamefont
  {Chakravarty}}, \bibinfo {author} {\bibfnamefont {A.~T.}\ \bibnamefont
  {Dorsey}}, \bibinfo {author} {\bibfnamefont {M.~P.~A.}\ \bibnamefont
  {Fisher}}, \bibinfo {author} {\bibfnamefont {A.}~\bibnamefont {Garg}}, \ and\
  \bibinfo {author} {\bibfnamefont {W.}~\bibnamefont {Zwerger}},\ }\href@noop
  {} {\bibfield  {journal} {\bibinfo  {journal} {Rev. Mod. Phys.},\ }\textbf
  {\bibinfo {volume} {59}},\ \bibinfo {pages} {1} (\bibinfo {year}
  {1987})}\BibitemShut {NoStop}%
\bibitem [{\citenamefont {Chang}\ and\ \citenamefont
  {Chakravarty}(1985)}]{Chang:1985}%
  \BibitemOpen
  \bibfield  {author} {\bibinfo {author} {\bibfnamefont {L.-D.}\ \bibnamefont
  {Chang}}\ and\ \bibinfo {author} {\bibfnamefont {S.}~\bibnamefont
  {Chakravarty}},\ }\href@noop {} {\bibfield  {journal} {\bibinfo  {journal}
  {Phys. Rev. B},\ }\textbf {\bibinfo {volume} {31}},\ \bibinfo {pages} {154}
  (\bibinfo {year} {1985})}\BibitemShut {NoStop}%
\bibitem [{\citenamefont {Chakravarty}\ and\ \citenamefont
  {Anderson}(1994)}]{Chakravarty:1994}%
  \BibitemOpen
  \bibfield  {author} {\bibinfo {author} {\bibfnamefont {S.}~\bibnamefont
  {Chakravarty}}\ and\ \bibinfo {author} {\bibfnamefont {P.~W.}\ \bibnamefont
  {Anderson}},\ }\href@noop {} {\bibfield  {journal} {\bibinfo  {journal}
  {Phys. Rev. Lett.},\ }\textbf {\bibinfo {volume} {72}},\ \bibinfo {pages}
  {3859} (\bibinfo {year} {1994})}\BibitemShut {NoStop}%
\bibitem [{\citenamefont {Ramshaw}\ \emph {et~al.}(2011)\citenamefont
  {Ramshaw}, \citenamefont {Vignolle}, \citenamefont {Liang}, \citenamefont
  {Hardy}, \citenamefont {Proust},\ and\ \citenamefont {Bonn}}]{Ramshaw:2011}%
  \BibitemOpen
  \bibfield  {author} {\bibinfo {author} {\bibfnamefont {B.~J.}\ \bibnamefont
  {Ramshaw}}, \bibinfo {author} {\bibfnamefont {B.}~\bibnamefont {Vignolle}},
  \bibinfo {author} {\bibfnamefont {R.}~\bibnamefont {Liang}}, \bibinfo
  {author} {\bibfnamefont {W.~N.}\ \bibnamefont {Hardy}}, \bibinfo {author}
  {\bibfnamefont {C.}~\bibnamefont {Proust}}, \ and\ \bibinfo {author}
  {\bibfnamefont {D.~A.}\ \bibnamefont {Bonn}},\ }\href@noop {} {\bibfield
  {journal} {\bibinfo  {journal} {Nat. Phys.},\ }\textbf {\bibinfo {volume}
  {7}},\ \bibinfo {pages} {234} (\bibinfo {year} {2011})}\BibitemShut {NoStop}%
\bibitem [{\citenamefont {Sebastian}\ \emph {et~al.}(2011)\citenamefont
  {Sebastian}, \citenamefont {Harrison}, \citenamefont {Altarawneh},
  \citenamefont {Balakirev}, \citenamefont {Mielke}, \citenamefont {Liang},
  \citenamefont {Bonn}, \citenamefont {Hardy},\ and\ \citenamefont
  {Lonzarich}}]{Sebastian:2011}%
  \BibitemOpen
  \bibfield  {author} {\bibinfo {author} {\bibfnamefont {S.~E.}\ \bibnamefont
  {Sebastian}}, \bibinfo {author} {\bibfnamefont {N.}~\bibnamefont {Harrison}},
  \bibinfo {author} {\bibfnamefont {M.~M.}\ \bibnamefont {Altarawneh}},
  \bibinfo {author} {\bibfnamefont {F.~F.}\ \bibnamefont {Balakirev}}, \bibinfo
  {author} {\bibfnamefont {C.~H.}\ \bibnamefont {Mielke}}, \bibinfo {author}
  {\bibfnamefont {R.}~\bibnamefont {Liang}}, \bibinfo {author} {\bibfnamefont
  {D.~A.}\ \bibnamefont {Bonn}}, \bibinfo {author} {\bibfnamefont {W.~N.}\
  \bibnamefont {Hardy}}, \ and\ \bibinfo {author} {\bibfnamefont {G.~G.}\
  \bibnamefont {Lonzarich}},\ }\href@noop {} {\enquote {\bibinfo {title}
  {Direct observation of multiple spin zeroes in the underdoped high
  temperature superconductor $\mathrm{YBa_{2}Cu_{3}O_{6+x}}$},}\ } (\bibinfo
  {year} {2011}),\ \Eprint {http://arxiv.org/abs/1103.4178v1 [cond-mat]}
  {arXiv:1103.4178v1 [cond-mat]} \BibitemShut {NoStop}%
\bibitem [{\citenamefont {Wu}\ \emph {et~al.}(2011)\citenamefont {Wu},
  \citenamefont {Mayaffre}, \citenamefont {Kramer}, \citenamefont {Horvatic},
  \citenamefont {Berthier}, \citenamefont {Hardy}, \citenamefont {Liang},
  \citenamefont {Bonn},\ and\ \citenamefont {Julien}}]{Julien:2011}%
  \BibitemOpen
  \bibfield  {author} {\bibinfo {author} {\bibfnamefont {T.}~\bibnamefont
  {Wu}}, \bibinfo {author} {\bibfnamefont {H.}~\bibnamefont {Mayaffre}},
  \bibinfo {author} {\bibfnamefont {S.}~\bibnamefont {Kramer}}, \bibinfo
  {author} {\bibfnamefont {M.}~\bibnamefont {Horvatic}}, \bibinfo {author}
  {\bibfnamefont {C.}~\bibnamefont {Berthier}}, \bibinfo {author}
  {\bibfnamefont {W.}~\bibnamefont {Hardy}}, \bibinfo {author} {\bibfnamefont
  {R.}~\bibnamefont {Liang}}, \bibinfo {author} {\bibfnamefont
  {D.}~\bibnamefont {Bonn}}, \ and\ \bibinfo {author} {\bibfnamefont {M.-H.}\
  \bibnamefont {Julien}},\ }\href@noop {} {\enquote {\bibinfo {title}
  {Magnetic-field-induced stripe order in the high temperature superconductor
  $\mathrm{YBa_{2}Cu_{3}O_{y}}$},}\ } (\bibinfo {year} {2011})\BibitemShut
  {NoStop}%
\bibitem [{\citenamefont {Zheng}\ \emph {et~al.}(1999)\citenamefont {Zheng},
  \citenamefont {Clark}, \citenamefont {Kitaoka}, \citenamefont {Asayama},
  \citenamefont {Kodama}, \citenamefont {Kuhns},\ and\ \citenamefont
  {Moulton}}]{Zheng:1999}%
  \BibitemOpen
  \bibfield  {author} {\bibinfo {author} {\bibfnamefont {G.-q.}\ \bibnamefont
  {Zheng}}, \bibinfo {author} {\bibfnamefont {W.~G.}\ \bibnamefont {Clark}},
  \bibinfo {author} {\bibfnamefont {Y.}~\bibnamefont {Kitaoka}}, \bibinfo
  {author} {\bibfnamefont {K.}~\bibnamefont {Asayama}}, \bibinfo {author}
  {\bibfnamefont {Y.}~\bibnamefont {Kodama}}, \bibinfo {author} {\bibfnamefont
  {P.}~\bibnamefont {Kuhns}}, \ and\ \bibinfo {author} {\bibfnamefont {W.~G.}\
  \bibnamefont {Moulton}},\ }\href@noop {} {\bibfield  {journal} {\bibinfo
  {journal} {Phys. Rev. B},\ }\textbf {\bibinfo {volume} {60}},\ \bibinfo
  {pages} {R9947} (\bibinfo {year} {1999})}\BibitemShut {NoStop}%
\bibitem [{\citenamefont {Yelland}\ \emph {et~al.}(2008)\citenamefont
  {Yelland}, \citenamefont {Singleton}, \citenamefont {Mielke}, \citenamefont
  {Harrison}, \citenamefont {Balakirev}, \citenamefont {Dabrowski},\ and\
  \citenamefont {Cooper}}]{Yelland:2008}%
  \BibitemOpen
  \bibfield  {author} {\bibinfo {author} {\bibfnamefont {E.~A.}\ \bibnamefont
  {Yelland}}, \bibinfo {author} {\bibfnamefont {J.}~\bibnamefont {Singleton}},
  \bibinfo {author} {\bibfnamefont {C.~H.}\ \bibnamefont {Mielke}}, \bibinfo
  {author} {\bibfnamefont {N.}~\bibnamefont {Harrison}}, \bibinfo {author}
  {\bibfnamefont {F.~F.}\ \bibnamefont {Balakirev}}, \bibinfo {author}
  {\bibfnamefont {B.}~\bibnamefont {Dabrowski}}, \ and\ \bibinfo {author}
  {\bibfnamefont {J.~R.}\ \bibnamefont {Cooper}},\ }\href@noop {} {\bibfield
  {journal} {\bibinfo  {journal} {Phys. Rev. Lett.},\ }\textbf {\bibinfo
  {volume} {100}},\ \bibinfo {pages} {047003} (\bibinfo {year}
  {2008})}\BibitemShut {NoStop}%
\bibitem [{\citenamefont {Bangura}\ \emph {et~al.}(2008)\citenamefont
  {Bangura}, \citenamefont {Fletcher}, \citenamefont {Carrington},
  \citenamefont {Levallois}, \citenamefont {Nardone}, \citenamefont {Vignolle},
  \citenamefont {Heard}, \citenamefont {Doiron-Leyraud}, \citenamefont
  {LeBoeuf}, \citenamefont {Taillefer}, \citenamefont {Adachi}, \citenamefont
  {Proust},\ and\ \citenamefont {Hussey}}]{Bangura:2008}%
  \BibitemOpen
  \bibfield  {author} {\bibinfo {author} {\bibfnamefont {A.~F.}\ \bibnamefont
  {Bangura}}, \bibinfo {author} {\bibfnamefont {J.~D.}\ \bibnamefont
  {Fletcher}}, \bibinfo {author} {\bibfnamefont {A.}~\bibnamefont
  {Carrington}}, \bibinfo {author} {\bibfnamefont {J.}~\bibnamefont
  {Levallois}}, \bibinfo {author} {\bibfnamefont {M.}~\bibnamefont {Nardone}},
  \bibinfo {author} {\bibfnamefont {B.}~\bibnamefont {Vignolle}}, \bibinfo
  {author} {\bibfnamefont {P.~J.}\ \bibnamefont {Heard}}, \bibinfo {author}
  {\bibfnamefont {N.}~\bibnamefont {Doiron-Leyraud}}, \bibinfo {author}
  {\bibfnamefont {D.}~\bibnamefont {LeBoeuf}}, \bibinfo {author} {\bibfnamefont
  {L.}~\bibnamefont {Taillefer}}, \bibinfo {author} {\bibfnamefont
  {S.}~\bibnamefont {Adachi}}, \bibinfo {author} {\bibfnamefont
  {C.}~\bibnamefont {Proust}}, \ and\ \bibinfo {author} {\bibfnamefont {N.~E.}\
  \bibnamefont {Hussey}},\ }\href@noop {} {\bibfield  {journal} {\bibinfo
  {journal} {Phys. Rev. Lett.},\ }\textbf {\bibinfo {volume} {100}},\ \bibinfo
  {pages} {047004} (\bibinfo {year} {2008})}\BibitemShut {NoStop}%
\bibitem [{\citenamefont {Kawasaki}\ \emph {et~al.}(2011)\citenamefont
  {Kawasaki}, \citenamefont {Lin}, \citenamefont {Kuhns}, \citenamefont
  {Reyes},\ and\ \citenamefont {Zheng}}]{Kawasaki:2011}%
  \BibitemOpen
  \bibfield  {author} {\bibinfo {author} {\bibfnamefont {S.}~\bibnamefont
  {Kawasaki}}, \bibinfo {author} {\bibfnamefont {C.}~\bibnamefont {Lin}},
  \bibinfo {author} {\bibfnamefont {P.~L.}\ \bibnamefont {Kuhns}}, \bibinfo
  {author} {\bibfnamefont {A.~P.}\ \bibnamefont {Reyes}}, \ and\ \bibinfo
  {author} {\bibfnamefont {G.-q.}\ \bibnamefont {Zheng}},\ }\href@noop {}
  {\bibfield  {journal} {\bibinfo  {journal} {Phys. Rev. Lett.},\ }\textbf
  {\bibinfo {volume} {105}},\ \bibinfo {pages} {137002} (\bibinfo {year}
  {2011})}\BibitemShut {NoStop}%
\bibitem [{\citenamefont {Brown}(2011)}]{Brown:2011}%
  \BibitemOpen
  \bibfield  {author} {\bibinfo {author} {\bibfnamefont {S.~E.}\ \bibnamefont
  {Brown}},\ }\href@noop {} {} (\bibinfo {year} {2011})\BibitemShut {NoStop}%
\bibitem [{\citenamefont {Motoyama}\ \emph {et~al.}(2007)\citenamefont
  {Motoyama}, \citenamefont {Yu}, \citenamefont {Vishik}, \citenamefont {Vajk},
  \citenamefont {Mang},\ and\ \citenamefont {Greven}}]{Motoyama:2007}%
  \BibitemOpen
  \bibfield  {author} {\bibinfo {author} {\bibfnamefont {E.~M.}\ \bibnamefont
  {Motoyama}}, \bibinfo {author} {\bibfnamefont {G.}~\bibnamefont {Yu}},
  \bibinfo {author} {\bibfnamefont {I.~M.}\ \bibnamefont {Vishik}}, \bibinfo
  {author} {\bibfnamefont {O.~P.}\ \bibnamefont {Vajk}}, \bibinfo {author}
  {\bibfnamefont {P.~K.}\ \bibnamefont {Mang}}, \ and\ \bibinfo {author}
  {\bibfnamefont {M.}~\bibnamefont {Greven}},\ }\href@noop {} {\bibfield
  {journal} {\bibinfo  {journal} {Nature},\ }\textbf {\bibinfo {volume}
  {445}},\ \bibinfo {pages} {186} (\bibinfo {year} {2007})}\BibitemShut
  {NoStop}%
\bibitem [{\citenamefont {Nguyen}\ and\ \citenamefont
  {Chakravarty}(2002)}]{Nguyen:2002}%
  \BibitemOpen
  \bibfield  {author} {\bibinfo {author} {\bibfnamefont {H.~K.}\ \bibnamefont
  {Nguyen}}\ and\ \bibinfo {author} {\bibfnamefont {S.}~\bibnamefont
  {Chakravarty}},\ }\href@noop {} {\bibfield  {journal} {\bibinfo  {journal}
  {Phys. Rev. B},\ }\textbf {\bibinfo {volume} {65}},\ \bibinfo {pages}
  {180519} (\bibinfo {year} {2002})}\BibitemShut {NoStop}%
\bibitem [{\citenamefont {Garcia-Aldea}\ and\ \citenamefont
  {Chakravarty}(2011)}]{Garcia-Aldea:2011}%
  \BibitemOpen
  \bibfield  {author} {\bibinfo {author} {\bibfnamefont {D.}~\bibnamefont
  {Garcia-Aldea}}\ and\ \bibinfo {author} {\bibfnamefont {S.}~\bibnamefont
  {Chakravarty}},\ }\href@noop {} {\bibfield  {journal} {\bibinfo  {journal}
  {Phys. Rev. B},\ }\textbf {\bibinfo {volume} {82}},\ \bibinfo {pages}
  {184526} (\bibinfo {year} {2011})}\BibitemShut {NoStop}%
\bibitem [{\citenamefont {Norman}\ and\ \citenamefont
  {Lin}(2011)}]{Norman:2011}%
  \BibitemOpen
  \bibfield  {author} {\bibinfo {author} {\bibfnamefont {M.~R.}\ \bibnamefont
  {Norman}}\ and\ \bibinfo {author} {\bibfnamefont {J.}~\bibnamefont {Lin}},\
  }\href@noop {} {\bibfield  {journal} {\bibinfo  {journal} {Phys. Rev. B},\
  }\textbf {\bibinfo {volume} {82}},\ \bibinfo {pages} {060509} (\bibinfo
  {year} {2011})}\BibitemShut {NoStop}%
\bibitem [{\citenamefont {Ramazashvili}(2011)}]{Ramazashvili:2011}%
  \BibitemOpen
  \bibfield  {author} {\bibinfo {author} {\bibfnamefont {R.}~\bibnamefont
  {Ramazashvili}},\ }\href@noop {} {\bibfield  {journal} {\bibinfo  {journal}
  {Phys. Rev. Lett.},\ }\textbf {\bibinfo {volume} {105}},\ \bibinfo {pages}
  {216404} (\bibinfo {year} {2011})}\BibitemShut {NoStop}%
\bibitem [{\citenamefont {Mook}\ \emph {et~al.}(2002)\citenamefont {Mook},
  \citenamefont {Dai}, \citenamefont {Hayden}, \citenamefont {Hiess},
  \citenamefont {Lynn}, \citenamefont {Lee},\ and\ \citenamefont
  {Do\v{g}an}}]{Mook:2002}%
  \BibitemOpen
  \bibfield  {author} {\bibinfo {author} {\bibfnamefont {H.~A.}\ \bibnamefont
  {Mook}}, \bibinfo {author} {\bibfnamefont {P.}~\bibnamefont {Dai}}, \bibinfo
  {author} {\bibfnamefont {S.~M.}\ \bibnamefont {Hayden}}, \bibinfo {author}
  {\bibfnamefont {A.}~\bibnamefont {Hiess}}, \bibinfo {author} {\bibfnamefont
  {J.~W.}\ \bibnamefont {Lynn}}, \bibinfo {author} {\bibfnamefont {S.~H.}\
  \bibnamefont {Lee}}, \ and\ \bibinfo {author} {\bibfnamefont
  {F.}~\bibnamefont {Do\v{g}an}},\ }\href@noop {} {\bibfield  {journal}
  {\bibinfo  {journal} {Phys. Rev. B},\ }\textbf {\bibinfo {volume} {66}},\
  \bibinfo {pages} {144513} (\bibinfo {year} {2002})}\BibitemShut {NoStop}%
\bibitem [{\citenamefont {Mook}\ \emph {et~al.}(2004)\citenamefont {Mook},
  \citenamefont {Dai}, \citenamefont {Hayden}, \citenamefont {Hiess},
  \citenamefont {Lee},\ and\ \citenamefont {Do\v{g}an}}]{Mook:2004}%
  \BibitemOpen
  \bibfield  {author} {\bibinfo {author} {\bibfnamefont {H.~A.}\ \bibnamefont
  {Mook}}, \bibinfo {author} {\bibfnamefont {P.}~\bibnamefont {Dai}}, \bibinfo
  {author} {\bibfnamefont {S.~M.}\ \bibnamefont {Hayden}}, \bibinfo {author}
  {\bibfnamefont {A.}~\bibnamefont {Hiess}}, \bibinfo {author} {\bibfnamefont
  {S.~H.}\ \bibnamefont {Lee}}, \ and\ \bibinfo {author} {\bibfnamefont
  {F.}~\bibnamefont {Do\v{g}an}},\ }\href@noop {} {\bibfield  {journal}
  {\bibinfo  {journal} {Phys. Rev. B},\ }\textbf {\bibinfo {volume} {69}},\
  \bibinfo {pages} {134509} (\bibinfo {year} {2004})}\BibitemShut {NoStop}%
\bibitem [{\citenamefont {Luttinger}(1960)}]{Luttinger:1960}%
  \BibitemOpen
  \bibfield  {author} {\bibinfo {author} {\bibfnamefont {J.~M.}\ \bibnamefont
  {Luttinger}},\ }\href@noop {} {\bibfield  {journal} {\bibinfo  {journal}
  {Phys. Rev.},\ }\textbf {\bibinfo {volume} {119}},\ \bibinfo {pages} {1153}
  (\bibinfo {year} {1960})}\BibitemShut {NoStop}%
\bibitem [{\citenamefont {Chubukov}\ and\ \citenamefont
  {Morr}(1997)}]{Chubukov:1997}%
  \BibitemOpen
  \bibfield  {author} {\bibinfo {author} {\bibfnamefont {A.~V.}\ \bibnamefont
  {Chubukov}}\ and\ \bibinfo {author} {\bibfnamefont {D.~K.}\ \bibnamefont
  {Morr}},\ }\href@noop {} {\bibfield  {journal} {\bibinfo  {journal} {Phys.
  Rep.},\ }\textbf {\bibinfo {volume} {288}},\ \bibinfo {pages} {355} (\bibinfo
  {year} {1997})}\BibitemShut {NoStop}%
\bibitem [{\citenamefont {Altshuler}\ \emph {et~al.}(1998)\citenamefont
  {Altshuler}, \citenamefont {Chubukov}, \citenamefont {Dashevskii},
  \citenamefont {Finkel'stein},\ and\ \citenamefont {Morr}}]{Altshuler:1998}%
  \BibitemOpen
  \bibfield  {author} {\bibinfo {author} {\bibfnamefont {B.~L.}\ \bibnamefont
  {Altshuler}}, \bibinfo {author} {\bibfnamefont {A.~V.}\ \bibnamefont
  {Chubukov}}, \bibinfo {author} {\bibfnamefont {A.}~\bibnamefont
  {Dashevskii}}, \bibinfo {author} {\bibfnamefont {A.~M.}\ \bibnamefont
  {Finkel'stein}}, \ and\ \bibinfo {author} {\bibfnamefont {D.~K.}\
  \bibnamefont {Morr}},\ }\href@noop {} {\bibfield  {journal} {\bibinfo
  {journal} {Europhys. Lett.},\ }\textbf {\bibinfo {volume} {41}},\ \bibinfo
  {pages} {401} (\bibinfo {year} {1998})}\BibitemShut {NoStop}%
\bibitem [{\citenamefont {Chakravarty}(2011)}]{Chakravarty:2011}%
  \BibitemOpen
  \bibfield  {author} {\bibinfo {author} {\bibfnamefont {S.}~\bibnamefont
  {Chakravarty}},\ }\href@noop {} {\bibfield  {journal} {\bibinfo  {journal}
  {Rep. Prog. Phys.},\ }\textbf {\bibinfo {volume} {74}},\ \bibinfo {pages}
  {022501} (\bibinfo {year} {2011})}\BibitemShut {NoStop}%
\bibitem [{\citenamefont {Casey}\ and\ \citenamefont
  {Anderson}(2011)}]{Casey:2011}%
  \BibitemOpen
  \bibfield  {author} {\bibinfo {author} {\bibfnamefont {P.~A.}\ \bibnamefont
  {Casey}}\ and\ \bibinfo {author} {\bibfnamefont {P.~W.}\ \bibnamefont
  {Anderson}},\ }\href@noop {} {\bibfield  {journal} {\bibinfo  {journal}
  {Phys. Rev. Lett.},\ }\textbf {\bibinfo {volume} {106}},\ \bibinfo {pages}
  {097002} (\bibinfo {year} {2011})}\BibitemShut {NoStop}%
\end{thebibliography}
\end{document}